\documentclass[a4paper,12pt]{article}

\usepackage{bbold}
\usepackage{float}
\usepackage{amsmath, amssymb, setspace,cite}
\usepackage{graphicx}
\usepackage{booktabs}
\usepackage{cite}
\usepackage{hyperref}

\usepackage{color}

\newcommand{\ra}[1]{\renewcommand{\arraystretch}{#1}}

\newcommand{\eq}[1]{\begin{equation} #1 \end{equation}}

\newcommand{\eqa}[1]{\begin{eqnarray} #1 \end{eqnarray}}
\newcommand{\nn}{\nonumber}

\newcommand{\fl}{F_L}
\newcommand{\bin}{{\rm bin}}
\newcommand{\av}[1]{\langle #1 \rangle}
\newcommand{\C}[1]{{\cal{C}}_{#1}}
\newcommand{\Cp}[1]{{\cal{C}}^\prime_{#1}}
\newcommand{\Cnp}[1]{{\delta {\cal{C}}_{#1}}}
\newcommand{\Csm}[1]{{\cal{C}}^{\rm SM}_{#1}}

\newcommand{\op}[1]{{\cal O}_{#1}}
\newcommand{\heff}{{\cal H}_{\rm eff}}
\newcommand{\eg}{\emph{e.g.}}

\newcommand{\SuperIso}[1]{{\tt SuperIso#1}}
\newcommand{\softsusy}[1]{{\tt SOFTSUSY#1}}

\vspace{0.6cm}
\date{\today} 

\begin{document}

\hfill {\tt CERN-PH-TH/2014-004}

\hfill {\tt QFET-2014-01}

\hfill {\tt SI-HEP-2014-03}

\def\thefootnote{\fnsymbol{footnote}}

\begin{center}
\Large\bf\boldmath
\vspace*{1.cm} 
$B\to K^*\mu^+\mu^-$ optimised observables in the MSSM\unboldmath
\end{center}
\vspace{0.6cm}

\begin{center}
F.~Mahmoudi$^{1,2,}$\footnote{Electronic address: mahmoudi@in2p3.fr}, 
S.~Neshatpour$^{1,3,}$\footnote{Electronic address: neshatpour@clermont.in2p3.fr},
J.~Virto$^{4,}$\footnote{Electronic address: virto@physik.uni-siegen.de}\\[0.4cm] 
\vspace{0.6cm}
{\sl $^1$ Clermont Universit{\'e}, Universit\'e Blaise Pascal, CNRS/IN2P3,\\
LPC, BP 10448, 63000 Clermont-Ferrand, France}\\[0.4cm]
{\sl $^2$ CERN Theory Division, Physics Department\\ CH-1211 Geneva 23, Switzerland}\\[0.4cm]
{\sl $^3$ Department of Physics, Isfahan University of Technology,\\
Isfahan 84156-83111, Iran}\\[0.4cm]
{\sl $^4$ 
Theoretische Physik 1, Naturwissenschaftlich-Technische Fakult\"at, \\
Universit\"at Siegen,  D-57068 Siegen, Germany}
\end{center}

\renewcommand{\thefootnote}{\arabic{footnote}}
\setcounter{footnote}{0}

\vspace{1.cm}
\begin{abstract}
We provide a detailed analysis of the impact of the newly measured optimised observables in the $B\to K^*\mu^+\mu^-$ decay by the LHCb experiment. The analysis is performed in the MSSM, both in the context of the usual constrained scenarios, and in the context of a more general set-up where the SUSY partner masses are independent. We show that the global agreement of the MSSM solutions with the data is still very good. Nevertheless, especially in the constrained scenarios, the limits from $B\to K^*\mu^+\mu^-$ are now very strong, and are at the same level as the well-known $b\to s \gamma$ constraints. We describe the implications of the $B\to K^*\mu^+\mu^-$ measurements both on the Wilson coefficients and on the SUSY parameters.
\end{abstract}

\newpage

%%%%%%%%%%%%%%%%%%%%%%%%%%%%%%%%%%%%%%%%%%%%%%%%%%%%%%%%
\section{Introduction}
Rare $b \to s$ transitions provide a powerful probe of the flavour sector of the Standard Model (SM).
The effective Hamiltonian formulation of these processes allow for a convenient low energy description in 
terms of short distance Wilson coefficients $\C{i}$.
Deviations from the SM predictions can then be consistently parametrised in terms of New Physics (NP) contributions to the Wilson coefficients ($\delta \C{i}\equiv \C{i}-\Csm{i}$). In addition, model predictions for the coefficients $\delta \C{i}$ can be computed by matching the model to the effective theory at the electroweak scale; thus these coefficients provide a link between the high energy model and the low energy phenomenology. In the case of $b\to s$ 
transitions, the relevant $d=6$ Hamiltonian is given by
\eq{\heff= \heff^{\rm sl} + \heff^{\rm had}\ ,}
where the semileptonic part $\heff^{\rm sl}$ is composed of radiative and dileptonic operators $[\bar s \Gamma^{\mu\nu} b] F_{\mu\nu}$ and $[\bar s \Gamma b][ \bar\ell \Gamma'\ell]$, and the hadronic part $\heff^{\rm had}$ contains chromomagnetic $[\bar s \Gamma^{\mu\nu} b] G_{\mu\nu}$ and 4-quark operators $[\bar s \Gamma b][ \bar q_1 \Gamma' q_2]$.

Recently, a significant amount of work has been devoted to the study of $b\to s\ell\ell$ processes \cite{1307.5683,1308.1501,1310.2478,1312.5267,1310.3887,1310.3722,1308.1959,1310.1082,1310.1937,1308.4379}, 
where experimental progress has been outstanding (see \emph{e.g.} Refs.~\cite{1304.6325,1308.1707,1307.5025,1307.5024,atlas-2013-038} ). These processes are the most sensitive to $\heff^{\rm sl}$, and together with $b\to s\gamma$ processes, pose strong conditions on the corresponding Wilson coefficients. More specifically, $\heff^{\rm sl}$ is given by
\begin{equation}
{\cal H}_{\rm eff}^{\rm sl}  =  -\frac{4G_{F}}{\sqrt{2}} V_{tb} V_{ts}^{*} \, \Bigl[\,
\sum_{i=7,9,10} \Bigl( {\cal C}_{i} {\cal O}_{i} + {\cal C}_{i}^{\prime} {\cal O}_{i}^{\prime} \Bigr)\;+
    \sum_{i=1,2} \Bigl( {\cal C}_{Q_i} Q_{i} + {\cal C}_{Q_i}^{\prime} Q_{i}^{\prime} \Bigr) \Bigr]\; .
\end{equation}
The various operators in ${\cal H}_{\rm eff}^{\rm sl}$ as well as the chromomagnetic operator ${\cal O}_8$ are defined as
\begin{align}
\label{physical_basis}
{\cal O}_7^{(\prime)} &= \frac{e}{(4\pi)^2} m_b [\bar{s} \sigma^{\mu\nu} P_{R(L)} b] F_{\mu\nu} \;, &
{\cal O}_8^{(\prime)} &= \frac{g}{(4\pi)^2} m_b [\bar{s} \sigma^{\mu \nu} T^a P_{R(L)} b] G_{\mu \nu}^a \;, 
\\ \nonumber
{\cal O}_9^{(\prime)} &=  \frac{e^2}{(4\pi)^2} [\bar{s} \gamma^\mu P_{L(R)} b] [\bar{\ell} \gamma_\mu \ell] \;, &
{\cal O}_{10}^{(\prime)} &=  \frac{e^2}{(4\pi)^2} [\bar{s} \gamma^\mu P_{L(R)} b] [\bar{\ell} \gamma_\mu \gamma_5 \ell]\;, \\ \nonumber
{Q}_1^{(\prime)} &= \frac{e^2}{(4\pi)^2} [\bar{s} P_{R(L)} b][\bar{\ell}\,\ell]\;, &
{Q}_2^{(\prime)} &=  \frac{e^2}{(4\pi)^2} [\bar{s} P_{R(L)} b][\bar{\ell}\gamma_5 \ell]\;,
\end{align}
where $Q_{1,2}$ are the scalar and pseudo-scalar operators, related to the usual $\op{S,P}^{(\prime)}$ by the running $b$-quark mass: $\op{S,P}^{(\prime)}=\hat m_b\,Q_{1,2}^{(\prime)}$ (see {\it e.g.} Ref.~\cite{0811.1214}).
We have neglected the tensor operators as they are highly suppressed both in the SM and in the Minimal Supersymmetric Standard Model (MSSM).
The corresponding Wilson coefficients are well known in the SM \cite{9910220,0312090,0401041,0411071,0504194,0512066,0612329}, 
where $\Csm{\rm 7}=-0.297$, $\Csm{\rm 8}=-0.161$, $\Csm9=4.22$ and $\Csm{10}=-4.15$ 
at the scale of the $b$-quark pole mass ($\mu=m_b^{\rm pole}$), while their primed counterparts as well as the scalar and pseudo-scalar Wilson coefficients vanish to a good approximation.
Global fits to all available  $b\to s\gamma$ and $b\to s\mu^+\mu^-$ data constrain significantly the allowed values for the NP contributions $\delta \C{i}$ \cite{1111.1257,1205.1838,1207.0688,1207.2753}, specially when the latest  $B\to K^*\mu^+\mu^-$ data is included \cite{1307.5683,1308.1501,1310.2478,1312.5267}.

It is precisely the decay $B\to K^*(\to K\pi)\mu^+\mu^-$ that has attracted most of the attention recently. Its angular distribution \cite{9907386} (see also \eg \cite{0811.1214,1212.2263}) provides a plethora of observables sensitive to different helicity structures in the decay amplitude. However, theoretical predictions for the most obvious observables inherit large uncertainties from not-so-well-known hadronic form factors. 
This has led to the construction of a number of \emph{optimised} observables as appropriate ratios of angular coefficients where most of the dependence on the form factors cancels, while having high sensitivity to NP effects~\cite{0502060,1106.3283,1202.4266,1207.2753,1006.5013,1212.2321}. A complete minimal set of such observables is given by the $\{P_i^{(\prime)}\}$ ensemble \cite{1202.4266,1207.2753}. Rigorous SM predictions for these and other $B\to K^*\mu^+\mu^-$ observables can be found in Ref.~\cite{1303.5794}, and are also implemented within \SuperIso~\cite{SuperIso}.

While all previous experimental results from rare $|\Delta B|= |\Delta S|=1$ processes 
were pointing to SM like values for the Wilson coefficients, the bounds on dileptonic operators were relatively mild still allowing $|\Cnp{9,10}^{(\prime)}/\Csm{9,10}{}^{(\prime)}|\sim \op{}(1)$ \cite{1207.2753}. 
First measurements of the full angular distribution by the LHCb collaboration \cite{1304.6325,1308.1707} have changed the situation, presenting a tension in several binned observables, such as $P_2$ and $P_5'$ .

In Ref.~\cite{1307.5683}, the first model independent global fit for the Wilson coefficients using the recent LHCb results on $B\to K^*\mu^+\mu^-$ 
combined with other existing $b\to s$ data was performed, hinting to a rather large negative NP contribution $\delta \C{9} \sim -1.2$.
This was followed by a global fit in Ref.~\cite{1308.1501} indicating a value close to $\delta \C{9} \sim -0.9$ (in agreement with Ref.~\cite{1307.5683}), and a global Bayesian fit in Ref.~\cite{1310.2478} with a best fit value for $\delta \C{9}$ of either $-1.3$ or $-0.3$ depending on the set of observables used.
An independent check has been given in Ref.~\cite{1310.3887}: a fit to $B\to K^*\mu^+\mu^-$ and $B_s\to\phi\mu^+\mu^-$ observables at low hadronic recoil only, based on the recent unquenched lattice results for the relevant form factors \cite{1310.3722}, leads to a similar value for $\delta \C{9}\sim -1$. 
In Ref.~\cite{1312.5267} it was shown that a consistent new physics explanation
of the $\bar B\to \bar K^*\mu^+\mu^-$ results in the context of minimal flavour violation is possible.

The model-independent global fits require $\C7$ and $\C{10}$ to be close to their SM values (in order to satisfy the strong $B\to X_s\gamma$ and $B_s\to\mu^+\mu^-$ constraints). 
 This has promoted NP scenarios where $\delta \C{7}^{(\prime)}$ and $\delta \C{10}^{(\prime)}$ 
are negligible while $\delta \C{9}\sim-1$, as is the case with models with a $Z^\prime$ gauge boson coupling to left- and right-handed leptons with equal strength~\cite{1211.1896,1307.5683,1308.1959} (arising for example in the 331 models \cite{1310.1082,1311.6729}) or scenarios with effective four-quark scalar interactions~\cite{1310.1937}. It has also been argued that within SUSY models or in models with partial compositeness, such large values for $\delta \C{9}$ can hardly be obtained \cite{1308.1501}. The analysis in Ref.~\cite{1312.5267} shows that the set of solutions  with flipped sign SM Wilson coefficients has only a slightly larger $\chi^2$, although this scenario is considerably more aggressive in terms of New Physics.

The $\bar B\to \bar K^*\mu^+\mu^-$ decay has already been used to constrain the MSSM parameters (see {\it e.g.} \cite{1205.1500,1205.1845}).
In this work we study the implications of recent LHCb results for the $\bar B\to \bar K^*\mu^+\mu^-$ decay (combined with existing experimental results for other $b \to s$ transitions) on the MSSM.
Our numerical analysis is performed with \softsusy~\cite{softsusy} and \SuperIso~\cite{SuperIso} and we study two classes of scenarios within the MSSM: the constrained MSSM models CMSSM and the Non Universal Higgs Mass (NUHM) model and a more general set-up known as phenomenological MSSM (pMSSM) which enables access to regions of the MSSM parameter space otherwise inaccessible with the aforementioned constrained scenarios.
To investigate the viability of the MSSM scenarios considering recent experimental results on $\bar B\to \bar K^*\mu^+\mu^-$ observables, we perform a model-dependent analysis on the Wilson coefficients using the other relevant $b \to s$ modes as well.
Using this global analysis, constraints are obtained for several SUSY parameters within each scenario.

This paper is organised as follows: in section~\ref{sec:2} we provide the SM predictions and errors for $\bar B\to \bar K^*\mu^+\mu^-$ observables as well as the other $b\to s$ processes considered in this work. In section~\ref{sec:MSSM} we present the parameter ranges within which the MSSM scenarios
that we have considered are investigated. Our statistical treatment is described in Section~\ref{sec:stat}, and Section~\ref{sec:results} contains our results on the global analysis.
Conclusions are given in section~\ref{sec:conclusions}.

%%%%%%%%%%%%%%%%%%%%%%%%%%%%%%%%%%%%%%%%%%%%%%%%%%%%%%%%
\section{Optimised observables}
\label{sec:2}

Angular observables in $\bar B\to \bar K^*(\to K\pi)\mu^+\mu^-$ arise from the differential decay rate:
\eqa{\label{angdist}
\frac{d^4\Gamma(\bar{B}_d)}{dq^2\,d\!\cos\theta_K\,d\!\cos\theta_l\,d\phi}&=&\frac9{32\pi} \bigg[
J_{1s} \sin^2\theta_K + J_{1c} \cos^2\theta_K + (J_{2s} \sin^2\theta_K + J_{2c} \cos^2\theta_K) \cos 2\theta_l\nn\\[1.5mm]
&&\hspace{-2.7cm}+ J_3 \sin^2\theta_K \sin^2\theta_l \cos 2\phi + J_4 \sin 2\theta_K \sin 2\theta_l \cos\phi  + J_5 \sin 2\theta_K \sin\theta_l \cos\phi \nn\\[1.5mm]
&&\hspace{-2.7cm}+ (J_{6s} \sin^2\theta_K +  {J_{6c} \cos^2\theta_K})  \cos\theta_l    
+ J_7 \sin 2\theta_K \sin\theta_l \sin\phi  + J_8 \sin 2\theta_K \sin 2\theta_l \sin\phi \nn\\[1.5mm]
&&\hspace{-2.7cm}+ J_9 \sin^2\theta_K \sin^2\theta_l \sin 2\phi \bigg]\,,
}
where notation and conventions follow Refs.~\cite{0811.1214,1006.5013,1202.4266}. Here we focus on CP-averaged quantities, for which the angular distribution $d\bar \Gamma$ of the CP-conjugated process $B\to K^*\mu^+\mu^-$ must also be considered. This is obtained from Eq.~(\ref{angdist}) by replacing $J_{1,2,3,4,7}\to \bar J_{1,2,3,4,7}$ and $J_{5,6,8,9}\to -\bar J_{5,6,8,9}$, where $\bar J$ is equal to $J$ with all weak phases conjugated  (see \emph{e.g.} Ref.~\cite{9907386}).

The basic observables are the functions $J_i(q^2)$ integrated in $q^2$-bins ($q^2$ is the squared invariant mass of the muon pair). From these a number of \emph{optimised} observables can be constructed, where some hadronic uncertainties are minimised by taking appropriate ratios. A set of such optimised observables have been measured recently by the LHCb collaboration \cite{1304.6325,1308.1707}, these are defined as \cite{1202.4266,1207.2753,1303.5794}:

\begin{align}
\av{P_1}_{\rm bin}&= \frac12 \frac{\int_{{\rm bin}} dq^2 [J_3+\bar J_3]}{\int_{{\rm bin}} dq^2 [J_{2s}+\bar J_{2s}]}\ ,
&\av{P_2}_{\rm bin} &= \frac18 \frac{\int_{{\rm bin}} dq^2 [J_{6s}+\bar J_{6s}]}{\int_{{\rm bin}} dq^2 [J_{2s}+\bar J_{2s}]}\ ,\nn\\
\av{P'_4}_{\rm bin} &= \frac1{{\cal N}_\bin^\prime} \int_{{\rm bin}} dq^2 [J_4+\bar J_4]\ ,
&\av{P'_5}_{\rm bin} &= \frac1{2{\cal N}_\bin^\prime} \int_{{\rm bin}} dq^2 [J_5+\bar J_5]\ ,\nn\\
\av{P'_6}_{\rm bin} &= \frac{-1}{2{\cal N}_\bin^\prime} \int_{{\rm bin}} dq^2 [J_7+\bar J_7]\ ,
& \av{P'_8}_{\rm bin} &= \frac{-1}{{\cal N}_\bin^\prime} \int_{{\rm bin}} dq^2 [J_8+\bar J_8]\ ,
\label{Pi}
\end{align}
where the normalisation ${\cal N}_\bin^\prime$ is given by
\eq{{\cal N}_\bin^\prime = {\textstyle \sqrt{-\int_\bin dq^2 [J_{2s}+\bar J_{2s}] \int_{{\rm bin}} dq^2 [J_{2c}+\bar J_{2c}]}}\ .}

In addition, several non-optimised observables can be introduced in the analysis which are theoretically independent from the previous ones. In this paper we will consider the branching ratio and the longitudinal polarisation fraction:
\begin{align}
\av{d{\rm BR}/dq^2}_{\rm bin}&=\tau_B \frac{\av{d \Gamma/dq^2}_{\rm bin}+\av{d \bar \Gamma/dq^2}_{\rm bin}}2\ ,
&\av{\fl}_{\rm bin}&=- \frac{\int_{\rm bin} dq^2 [J_{2c}+\bar J_{2c}]}{\av{d \Gamma/dq^2}+\av{d \bar \Gamma/dq^2}} \ ,
\label{acp}
\end{align}
where
\eq{
\av{d \Gamma/dq^2}_{\rm bin} =\frac14 \int_{\rm bin} dq^2 [3 J_{1c} + 6 J_{1s} - J_{2c} -2 J_{2s}]
}
and analogously for $\bar\Gamma$. We do not consider the forward-backward asymmetry, which is significantly correlated to the observable $P_2$~\cite{1312.5267}. The set of SM predictions and experimental measurements for all these observables are summarised in Table~\ref{tab:obs}.

\begin{table}
 \ra{1.05}
\begin{center}
\footnotesize{\begin{tabular}{@{}lccr@{}}\hline
Observable & Experiment & SM prediction & Pull \\
\hline
%%%%%%%%%%%%%%%%%%%%%%%%%%%%%%%%%%%%%%%%%%%%%%%%%%%%%%%%%%%%%%%
%%%%%%%%%%%%%%%%%%%%%%%%%%%%%%%%%%%%%%%%%%%%%%%%%%%%%%%%%%%%%%
$10^4 BR(B \to X_s \gamma)$ & $3.43 \pm 0.22$ & $3.09 \pm 0.24$ & $+1.0$\\
$10^2 \Delta_0(B \to K^* \gamma)$ & $5.2 \pm 2.6$  & $7.9 \pm 3.9$ & $-0.6$\\
$10^9 BR(B_s \to \mu^+\mu^-)$ & $2.9 \pm 0.7$ & $3.49 \pm 0.38$ & $-0.7$\\
$10^6 BR(B \to X_s \mu^+\mu^-)_{q^2\in[1,6] \rm{GeV}^2}$ & $1.60 \pm 0.68$ & $1.73 \pm 0.16$ & $-0.2$\\
$10^6 BR(B \to X_s \mu^+\mu^-)_{q^2>14.4 \rm{GeV}^2}$ & $0.42 \pm 0.14$ &  $0.22 \pm 0.04$ & $+1.4$\\
\hline
%%%%%%%%%%%%%%%%%%%%%%%%%%%%%%%%%%%%%%%%%%%%%%%%%%%%%%%%%%%%%%%
%%%%%%%%%%%%%%%%%%%%%%%%%%%%%%%%%%%%%%%%%%%%%%%%%%%%%%%%%%%%%%
$10^7\langle dBR/dq^2(B \to K^* \mu^+\mu^-) \rangle_{q^2\in[0.1,2] \rm{GeV}^2}$ & $0.60 \pm 0.10$ & $0.70 \pm 0.81$ & $-0.1$\\
$\langle F_{L}(B \to K^* \mu^+ \mu^-) \rangle_{q^2\in[0.1,2] \rm{GeV}^2}$ & $ 0.37 \pm 0.11$ & $ 0.32 \pm 0.20 $ & $+0.2$\\
$\langle P_1(B \to K^* \mu^+\mu^-) \rangle_{q^2\in[0.1,2] \rm{GeV}^2}$ & $ -0.19 \pm 0.40 $ & $ -0.01 \pm 0.04 $ & $-0.4$ \\
$\langle P_2(B \to K^* \mu^+ \mu^-) \rangle_{q^2\in[0.1,2] \rm{GeV}^2}$ & $ 0.03 \pm 0.15$ & $ 0.17 \pm 0.02 $ & $-0.9$ \\
$\langle P_4'(B \to K^* \mu^+ \mu^-) \rangle_{q^2\in[0.1,2] \rm{GeV}^2}$ & $0.00 \pm 0.52$ & $ -0.37 \pm 0.03 $ & $+0.7$ \\
$\langle P_5'(B \to K^* \mu^+ \mu^-) \rangle_{q^2\in[0.1,2] \rm{GeV}^2}$ & $0.45 \pm 0.24$ & $ 0.52 \pm 0.04 $ &  $-0.3$\\
$\langle P_6'(B \to K^* \mu^+ \mu^-) \rangle_{q^2\in[0.1,2] \rm{GeV}^2}$ & $0.24 \pm 0.23$ & $ -0.05 \pm 0.04 $ & $+1.3$ \\
$\langle P_8'(B \to K^* \mu^+ \mu^-) \rangle_{q^2\in[0.1,2] \rm{GeV}^2}$ & $-0.12 \pm 0.56$ & $ 0.02 \pm 0.04 $ & $-0.2$ \\
\hline
%%%%%%%%%%%%%%%%%%%%%%%%%%%%%%%%%%%%%%%%%%%%%%%%%%%%%%%%%%%%%%%
%%%%%%%%%%%%%%%%%%%%%%%%%%%%%%%%%%%%%%%%%%%%%%%%%%%%%%%%%%%%%%
$10^7 \langle dBR/dq^2(B \to K^* \mu^+\mu^-) \rangle_{q^2\in[2,4.3] \rm{GeV}^2}$ & $0.30 \pm 0.05$ & $ 0.35 \pm 0.29 $ & $-0.2$ \\
$\langle F_{L}(B \to K^* \mu^+ \mu^-) \rangle_{q^2\in[2,4.3] \rm{GeV}^2}$ & $ 0.74 \pm 0.10$ & $ 0.76 \pm 0.20$ & $-0.1$ \\
$\langle P_1(B \to K^* \mu^+\mu^-) \rangle_{q^2\in[2,4.3] \rm{GeV}^2}$ & $ -0.29 \pm 0.65$ & $ -0.05 \pm 0.05 $ & $-0.4$ \\
$\langle P_2(B \to K^* \mu^+\mu^-) \rangle_{q^2\in[2,4.3] \rm{GeV}^2}$ & $ 0.50 \pm 0.08 $ & $ 0.25 \pm 0.09 $ & $+2.0$ \\
$\langle P_4'(B \to K^* \mu^+\mu^-) \rangle_{q^2\in[2,4.3] \rm{GeV}^2}$ & $0.74 \pm 0.60$ & $ 0.54 \pm 0.07 $ & $+0.3$ \\
$\langle P_5'(B \to K^* \mu^+\mu^-) \rangle_{q^2\in[2,4.3] \rm{GeV}^2}$ & $0.29 \pm 0.40$ & $ -0.33 \pm 0.11 $ &  $+1.5$\\
$\langle P_6'(B \to K^* \mu^+\mu^-) \rangle_{q^2\in[2,4.3] \rm{GeV}^2}$ & $-0.15 \pm 0.38$ & $ -0.06 \pm 0.06 $ & $-0.2$ \\
$\langle P_8'(B \to K^* \mu^+\mu^-) \rangle_{q^2\in[2,4.3] \rm{GeV}^2}$ & $-0.3 \pm 0.60$ & $ 0.04 \pm 0.05 $ & $-0.6$ \\
\hline
%%%%%%%%%%%%%%%%%%%%%%%%%%%%%%%%%%%%%%%%%%%%%%%%%%%%%%%%%%%%%%%
%%%%%%%%%%%%%%%%%%%%%%%%%%%%%%%%%%%%%%%%%%%%%%%%%%%%%%%%%%%%%%
$10^7 \langle dBR/dq^2(B \to K^* \mu^+\mu^-) \rangle_{q^2\in[4.3,8.68] \rm{GeV}^2}$ & $0.49 \pm 0.08$ & $0.48 \pm 0.53$ & $+0.0$ \\
$\langle F_{L}(B \to K^* \mu^+ \mu^-) \rangle_{q^2\in[4.3,8.68] \rm{GeV}^2}$ & $ 0.57 \pm 0.08$ & $ 0.63 \pm 0.14$ & $-0.4$ \\
$\langle P_1(B \to K^* \mu^+\mu^-) \rangle_{q^2\in[4.3,8.68] \rm{GeV}^2}$ & $ 0.36 \pm 0.31$ & $ -0.11 \pm 0.06 $ & $+1.5$ \\
$\langle P_2(B \to K^* \mu^+\mu^-) \rangle_{q^2\in[4.3,8.68] \rm{GeV}^2}$ & $ -0.25 \pm 0.08$ & $ -0.36 \pm 0.05$ & $+1.1$ \\
$\langle P_4'(B \to K^* \mu^+\mu^-) \rangle_{q^2\in[4.3,8.68] \rm{GeV}^2}$ & $1.18 \pm 0.32$ & $ 0.99 \pm 0.03 $ & $+0.6$ \\
$\langle P_5'(B \to K^* \mu^+\mu^-) \rangle_{q^2\in[4.3,8.68] \rm{GeV}^2}$ & $-0.19 \pm 0.16$ & $ -0.83 \pm 0.05 $ & $+3.8$ \\
$\langle P_6'(B \to K^* \mu^+\mu^-) \rangle_{q^2\in[4.3,8.68] \rm{GeV}^2}$ & $0.04 \pm 0.16$ & $ -0.02 \pm 0.06 $ & $+0.4$ \\
$\langle P_8'(B \to K^* \mu^+\mu^-) \rangle_{q^2\in[4.3,8.68] \rm{GeV}^2}$ & $0.58 \pm 0.38$ & $ 0.02 \pm 0.06 $ & $+1.4$ \\
\hline
%%%%%%%%%%%%%%%%%%%%%%%%%%%%%%%%%%%%%%%%%%%%%%%%%%%%%%%%%%%%%%%
%%%%%%%%%%%%%%%%%%%%%%%%%%%%%%%%%%%%%%%%%%%%%%%%%%%%%%%%%%%%%%
$10^7\langle dBR/dq^2(B \to K^* \mu^+\mu^-) \rangle_{q^2\in[14.18,16] \rm{GeV}^2}$ & $0.56 \pm 0.10$ & $0.67 \pm 1.17$ &  $-0.1$\\
$\langle F_{L}(B \to K^* \mu^+ \mu^-) \rangle_{q^2\in[14.18,16] \rm{GeV}^2}$ & $ 0.33 \pm 0.09$ & $ 0.39 \pm 0.24 $ & $-0.2$ \\
$\langle P_1(B \to K^* \mu^+\mu^-) \rangle_{q^2\in[14.18,16] \rm{GeV}^2}$ & $ 0.07 \pm 0.28$ & $ -0.32 \pm 0.70$ & $+0.5$ \\
$\langle P_2(B \to K^* \mu^+\mu^-) \rangle_{q^2\in[14.18,16] \rm{GeV}^2}$ & $ -0.50 \pm 0.03$ & $ -0.47 \pm 0.14$ & $-0.2$ \\
$\langle P_4'(B \to K^* \mu^+\mu^-) \rangle_{q^2\in[14.18,16] \rm{GeV}^2}$ & $-0.18 \pm 0.70$ & $ 1.15 \pm 0.33$ & $-1.7$ \\
$\langle P_5'(B \to K^* \mu^+\mu^-) \rangle_{q^2\in[14.18,16] \rm{GeV}^2}$ & $-0.79 \pm 0.27$ & $ -0.82 \pm 0.36$ & $+0.1$ \\
$\langle P_6'(B \to K^* \mu^+\mu^-) \rangle_{q^2\in[14.18,16] \rm{GeV}^2}$ & $0.18 \pm 0.25$ & $ 0.00 \pm 0.00$ & $+0.7$ \\
$\langle P_8'(B \to K^* \mu^+\mu^-) \rangle_{q^2\in[14.18,16] \rm{GeV}^2}$ & $-0.40 \pm 0.60$ & $ 0.00 \pm 0.01$ & $-0.7$ \\
\hline
%%%%%%%%%%%%%%%%%%%%%%%%%%%%%%%%%%%%%%%%%%%%%%%%%%%%%%%%%%%%%%%
%%%%%%%%%%%%%%%%%%%%%%%%%%%%%%%%%%%%%%%%%%%%%%%%%%%%%%%%%%%%%%
$10^7\langle dBR/dq^2(B \to K^* \mu^+\mu^-) \rangle_{q^2\in[16,19] \rm{GeV}^2}$ & $0.41 \pm 0.07$ & $0.43 \pm 0.78$ & $+0.0$ \\
$\langle F_{L}(B \to K^* \mu^+ \mu^-) \rangle_{q^2\in[16,19] \rm{GeV}^2}$ & $ 0.38 \pm 0.09$ & $ 0.36 \pm 0.13 $ & $+0.1$ \\
$\langle P_1(B \to K^* \mu^+\mu^-) \rangle_{q^2\in[16,19] \rm{GeV}^2}$ & $ -0.71 \pm 0.36$ & $ -0.55 \pm 0.59 $ & $-0.2$ \\
$\langle P_2(B \to K^* \mu^+\mu^-) \rangle_{q^2\in[16,19] \rm{GeV}^2}$ & $ -0.32 \pm 0.08$ & $ -0.41 \pm 0.15 $ & $+0.5$ \\
$\langle P_4'(B \to K^* \mu^+\mu^-) \rangle_{q^2\in[16,19] \rm{GeV}^2}$ & $0.70 \pm 0.52$ & $ 1.24 \pm 0.25 $ & $-0.9$ \\
$\langle P_5'(B \to K^* \mu^+\mu^-) \rangle_{q^2\in[16,19] \rm{GeV}^2}$ & $-0.60 \pm 0.21$ & $ -0.66 \pm 0.37 $ & $+0.1$ \\
$\langle P_6'(B \to K^* \mu^+\mu^-) \rangle_{q^2\in[16,19] \rm{GeV}^2}$ & $-0.31 \pm 0.39$ & $ 0.00 \pm 0.00 $ & $-0.8$ \\
$\langle P_8'(B \to K^* \mu^+\mu^-) \rangle_{q^2\in[16,19] \rm{GeV}^2}$ & $0.12 \pm 0.54$ & $ 0.00 \pm 0.04 $ & $+0.2$ \\
\hline
%%%%%%%%%%%%%%%%%%%%%%%%%%%%%%%%%%%%%%%%%%%%%%%%%%%%%%%%%%%%%%%
%%%%%%%%%%%%%%%%%%%%%%%%%%%%%%%%%%%%%%%%%%%%%%%%%%%%%%%%%%%%%%
% $\langle dBR/dq^2(B \to K^* \mu^+\mu^-) \rangle_{q^2\in[1,6] \rm{GeV}^2}$ & $(0.34 \pm 0.03 \pm 0.04 \pm 0.02 \pm 0.03)\times 10^{-7}$ & $(0.38 \pm 0.33)\times 10^{-7}$ & \\
% $\langle F_{L}(B \to K^* \mu^+ \mu^-) \rangle_{q^2\in[1,6] \rm{GeV}^2}$ & $ 0.65 \pm 0.08 \pm 0.03$ & $ 0.70 \pm 0.21 $ & \\
% $\langle P_1(B \to K^* \mu^+\mu^-) \rangle_{q^2\in[1,6] \rm{GeV}^2}$ & $ 0.15 \pm 0.41 \pm 0.03$ & $ -0.06 \pm 0.04 $ & \\
% $\langle P_2(B \to K^* \mu^+\mu^-) \rangle_{q^2\in[1,6] \rm{GeV}^2}$ & $ 0.33 \pm 0.12 \pm 0.02$ & $ 0.10 \pm 0.08 $ & \\
% $\langle P_4'(B \to K^* \mu^+\mu^-) \rangle_{q^2\in[1,6] \rm{GeV}^2}$ & $0.58 \pm 0.36 \pm 0.06$ & $ 0.53 \pm 0.07 $ & \\
% $\langle P_5'(B \to K^* \mu^+\mu^-) \rangle_{q^2\in[1,6] \rm{GeV}^2}$ & $0.21 \pm 0.21 \pm 0.03$ & $ -0.34 \pm 0.10 $ & \\
% $\langle P_6'(B \to K^* \mu^+\mu^-) \rangle_{q^2\in[1,6] \rm{GeV}^2}$ & $0.18 \pm 0.21 \pm 0.03$ & $ -0.05 \pm 0.05 $ & \\
% $\langle P_8'(B \to K^* \mu^+\mu^-) \rangle_{q^2\in[1,6] \rm{GeV}^2}$ & $0.46 \pm 0.38 \pm 0.04$ & $ 0.03 \pm 0.04$ & \\
% \hline
\end{tabular}}
\caption{The most recent experimental values and SM predictions for the observables used in this study. Experimental error bars are symmetrised by taking the largest sided uncertainty to both sides. Pulls are therefore slightly different with respect to Ref.~\cite{1307.5683}. The several sources of uncertainty are added in quadrature.
\label{tab:obs}}
\end{center}
\end{table}
For the computation of these $B\to K^*\mu^+\mu^-$ observables in terms of the Wilson coefficients we follow Ref.~\cite{1303.5794}. At large recoil (low $q^2$) the theoretical tools are those of QCDF/SCET as described in Refs.~\cite{0106067,0412400}. In this kinematical region several form-factor relations \cite{9812358,0008255} allow to build the various optimised observables \cite{1202.4266,1207.2753,1303.5794}. At low recoil (large $q^2$) the computation relies on an OPE for the relevant non-local hadronic matrix element, either in full QCD \cite{1101.5118} or within the HQET \cite{0404250}. Form-factor relations arise in HQET, allowing to construct optimised observables in this kinematic region \cite{1006.5013,1212.2321,1303.5794}, but the observables in Eq.~(\ref{Pi}) are not optimised at low recoil. The full form factors needed as an input are taken from the large-recoil LCSR computation in Ref.~\cite{1006.4945}. These are extrapolated to the low recoil region following the procedure detailed in 
Ref.~\cite{1303.5794}, and are consistent with lattice QCD results in this region \cite{0611295,1310.3722}, as well as with HQET form factor relations \cite{0404250,1303.5794}, for which a $20\%$ error from $1/m_b$ power corrections is included. Concerning power corrections to the hadronic contribution at low recoil, a $10\%$ error is added to each amplitude in an uncorrelated manner and with an arbitrary phase. The computation of these observables is implemented within \SuperIso~\cite{SuperIso}, and marginal numerical differences on the SM predictions with respect to Ref.~\cite{1303.5794} are due to slightly different choice of central values for input parameters.

A visualisation of how these observables depend on the various Wilson coefficients can be obtained by linearising with respect to the NP contributions to the coefficients. In addition, New Physics contributions to the primed coefficients $\Cp{7,9,10}$ are always negligible in the scenarios considered in this paper, being suppressed by a factor $m_s/m_b$ like in the SM. Within these approximations, we have:
\begin{equation}
\begin{array}{rcllll}
\delta \av{P_2}_{[0.1,2]} & \simeq  & + 0.37\,\delta \C{7} & + 0.02\,\delta \C{8}&                 & - 0.03\,\delta \C{10}  \\[2mm]
\delta \av{P_2}_{[2,4.3]} & \simeq  & - 2.48\,\delta \C{7} & - 0.10\,\delta \C{8}& - 0.17\,\delta \C{9} & + 0.03\,\delta \C{10} \\[2mm]
\delta \av{P_2}_{[4.3,8.68]} & \simeq  & - 0.71\,\delta \C{7} & - 0.04\,\delta \C{8}& - 0.09\,\delta \C{9} & - 0.04\,\delta \C{10} \\[4mm]
\delta \av{P'_4}_{[0.1,2]} & \simeq & + 0.59\,\delta \C{7} & & - 0.08\,\delta \C{9} & - 0.13\,\delta \C{10} \\[2mm]
\delta \av{P'_4}_{[2,4.3]} & \simeq  & + 2.45\,\delta \C{7} & + 0.11\,\delta \C{8}& + 0.06\,\delta \C{9} & - 0.14\,\delta \C{10} \\[2mm]
\delta \av{P'_4}_{[4.3,8.68]} & \simeq  & + 0.33\,\delta \C{7} & + 0.01\,\delta \C{8}& + 0.01\,\delta \C{9} & \\[4mm]
\delta \av{P'_5}_{[0.1,2]} & \simeq  & - 0.91\,\delta \C{7} & - 0.04\,\delta \C{8}& - 0.12\,\delta \C{9} & - 0.03\,\delta \C{10}\\[2mm]
\delta \av{P'_5}_{[2,4.3]} & \simeq  & - 3.04\,\delta \C{7} & - 0.14\,\delta \C{8} & - 0.29\,\delta \C{9} & - 0.03\,\delta \C{10}\\[2mm]
\delta \av{P'_5}_{[4.3,8.68]} & \simeq  & - 0.52\,\delta \C{7} & - 0.03\,\delta \C{8}& - 0.08\,\delta \C{9} & - 0.03\,\delta \C{10}
\end{array}
\label{SNexpr}
\end{equation}
while the rest of the observables are less sensitive to real NP contributions in $\C{7,8,9,10}$:
\begin{equation}
\begin{array}{rcllll}
\delta \av{P'_6}_{[0.1,2]} & \simeq  & - 0.21\,\delta \C{7} & + 0.09\,\delta \C{8}&  &   \\[2mm]
\delta \av{P'_6}_{[2,4.3]} & \simeq  & - 0.07\,\delta \C{7} & + 0.09\,\delta \C{8}&  &  \\[2mm]
\delta \av{P'_6}_{[4.3,8.68]} & \simeq  & + 0.02\,\delta \C{7} & + 0.04\,\delta \C{8}&  &  \\[4mm]
\delta \av{P'_8}_{[0.1,2]} & \simeq & + 0.20\,\delta \C{7} & - 0.11\,\delta \C{8} & + 0.01\,\delta \C{9} & \\[2mm]
\delta \av{P'_8}_{[2,4.3]} & \simeq  & + 0.11\,\delta \C{7} & - 0.10\,\delta \C{8}& + 0.01\,\delta \C{9} & + 0.02\,\delta \C{10} \\[2mm]
\delta \av{P'_8}_{[4.3,8.68]} & \simeq  & + 0.01\,\delta \C{7} & - 0.04\,\delta \C{8}&  & \\[4mm]
\end{array}
\label{SNexpr2}
\end{equation}
and $P_1$ depends (almost) only on right-handed currents. In addition, the observables $P'_5$ and $P'_6$ depend formally on scalar operators through the combination $(\C{Q_1}-\Cp{Q_1})$. However the coefficient in the linearised expression is smaller than 0.01 and has been neglected. We remind that the coefficients $\Cnp{i}$ are defined at the scale $m_b$.

The strongest constraint on $\delta \C{7}$ is provided by the branching ratio BR$(B\to X_s\gamma)$, which strongly favours values around $\delta \C{7}\in(-0.07,0.04)$, with preference for negative values. Looking at Eqs.~(\ref{SNexpr}) one can see that such negative small contributions to $\C{7}$ tend to increase the theory predictions for $\av{P_2}_{[2,4.3]}$, $\av{P_2}_{[4.3,8.68]}$, $\av{P'_5}_{[2,4.3]}$, $\av{P'_5}_{[4.3,8.68]}$ and decrease the value of $\av{P_2}_{[0.1,2]}$, improving the agreement with experiment. 
On the other hand the contributions to $\av{P'_5}_{[0.1,2]}$ and $\av{P'_4}$ go in the opposite direction, although still compatible within errors (experimental uncertainties on $P'_4$ are still rather large). Therefore, these observables will generally drift $\delta \C{7}$ towards small negative values in agreement with the expectations from BR$(B\to X_s\gamma)$ (see Ref.~\cite{1307.5683} and Fig.~5 therein).

However, in order to explain the observable $\av{P_5'}_{[4.3,8.68]}$, and to a lesser extent also $\av{P_2}_{[4.3,8.68]}$ and $\av{P_5'}_{[2,4.3]}$, contributions to other Wilson coefficients are necessary. As can be seen from Eqs.~(\ref{SNexpr}), the sensitivity to $\delta \C{10}$ is lower than to $\delta \C9$, so the latter is the best candidate. 
In addition,  $\delta \C{10}$ is constrained by BR$(B_s\to \mu^+\mu^-)$ to an extent that depends on the pattern of scalar contributions.
In the MSSM (where ${\cal C}_{Q_1}\approx -{\cal C}_{Q_2}$) this constraint is very effective.
In this context it was shown in Ref.~\cite{1307.5683} that the most favourable scenario involves a significant negative contribution to $\C9$. 

Such large NP contributions to $\C9$ are not possible in general within the MSSM (see also Ref.~\cite{1308.1501}), and as we shall see, moderately large negative values for $\delta \C9$ -- achievable within corners of the pMSSM -- are typically correlated to large values of other coefficients that lead to tensions with other flavour data. However, we will also see that all $B\to K^*\mu^+\mu^-$ observables including $\av{P_2}_{[4.3,8.68]}$ and $\av{P_5'}_{[2,4.3]}$ can be well described within the pMSSM, except for $\av{P_5'}_{[4.3,8.68]}$, which remains a challenge.

%%%%%%%%%%%%%%%%%%%%%%%%%%%%%%%%%%%%%%%%%%%%%%%%%%%%%%%%
\section{MSSM scenarios}
\label{sec:MSSM}

%%%%%%%%%%%%%%%%%
\begin{table}[!h]
\begin{center}
\begin{tabular}{|c|c|}
\hline
CMSSM Parameter  & ~~~~Range~~~~ \\
\hline\hline
$\tan\beta$ & [2, 60]\\
\hline
$m_0$ & [50, 3000] \\
\hline
$m_{1/2}$ & [50, 3000] \\
\hline
$A_0$ & [-10000, 10000] \\
\hline
${\rm sgn}(\mu)$ & $\pm$ \\
\hline
\multicolumn{1}{c}{~}\\
\multicolumn{1}{c}{~}\\
\multicolumn{1}{c}{~}\\
\hline
NUHM Parameter  & ~~~~Range~~~~ \\
\hline\hline
$\tan\beta$ & [2, 60]\\
\hline
$m_0$ & [50, 3000] \\
\hline
$m_{1/2}$ & [50, 3000] \\
\hline
$m_A$ & [50, 3000] \\
\hline
$A_0$ & [-10000, 10000] \\
\hline
$\mu$ & [-5000, 5000] \\
\hline
\end{tabular}~\quad\quad\begin{tabular}{|c|c|}
\hline
pMSSM Parameter & ~~~~Range~~~~ \\
\hline\hline
$\tan\beta$ & [2, 60]\\
\hline
$M_A$ & [50, 3000] \\
\hline
$M_1$ & [50, 3000] \\
\hline
$M_2$ & [50, 3000] \\
\hline
$M_3$ & [50, 3000] \\
\hline
$A_d=A_s=A_b$ & [-10000, 10000] \\
\hline
$A_u=A_c=A_t$ & [-10000, 10000] \\
\hline
$A_e=A_\mu=A_\tau$ & [-10000, 10000] \\
\hline
$\mu$ & [-5000, 5000] \\
\hline
$M_{\tilde{e}_L}=M_{\tilde{\mu}_L}$ & [50, 3000] \\
\hline
$M_{\tilde{e}_R}=M_{\tilde{\mu}_R}$ & [50, 3000] \\
\hline
$M_{\tilde{\tau}_L}$ & [50, 3000] \\
\hline
$M_{\tilde{\tau}_R}$ & [50, 3000] \\
\hline
$M_{\tilde{q}_{1L}}=M_{\tilde{q}_{2L}}$ & [50, 3000] \\
\hline
$M_{\tilde{q}_{3L}}$ & [50, 3000] \\
\hline
$M_{\tilde{u}_R}=M_{\tilde{c}_R}$ & [50, 3000] \\
\hline
$M_{\tilde{t}_R}$ & [50, 3000] \\
\hline
$M_{\tilde{d}_R}=M_{\tilde{s}_R}$ & [50, 3000] \\
\hline
$M_{\tilde{b}_R}$ & [50, 3000] \\
\hline
\end{tabular}
\end{center}
\caption{Parameter ranges adopted in the scans (in GeV when applicable) for the CMSSM, NUHM and pMSSM scenarios.\label{tab:ranges}}
\end{table}
%%%%%%%%%%%%%%%%%
%
We consider two classes of scenarios: the constrained MSSM scenarios CMSSM and NUHM, and a more general set-up with no universality assumptions for the sparticle masses at a high scale: the phenomenological MSSM (pMSSM)~\cite{9901246}.
We scan over the parameters of the MSSM scenarios using \softsusy{~3.3.10}~\cite{softsusy} in the ranges given in Table~\ref{tab:ranges}, and generate several millions of random points for each scenario. For each generated point, we then calculate the flavour observables with \SuperIso{~v3.4}~\cite{SuperIso}.
We apply a few loose cuts for some of the SUSY masses to avoid regions of the parameters that are excluded by direct SUSY searches. In particular we impose the gluino and squark masses to be above 500 GeV, and the chargino masses above 150 GeV. These limits could be much stronger in the CMSSM, but since the strong CMSSM limits are falsified in the case of pMSSM (see Ref.~\cite{1211.4004}) we consider the same loose cuts for all the scenarios we analyse here.

%%%%%%%%%%%%%%%%%%%%%%%%%%%%%%%%%%%%%%%%%%%%%%%%%%%%%%%%
\section{Statistical method}
\label{sec:stat}

In order to study the overall compatibility of the MSSM scenarios with the $B\to K^*\mu^+\mu^-$ measurements, including all other relevant flavour constraints, we perform a statistical $\chi^2$ analysis of the data. We consider all 45 observables collected in Table~\ref{tab:obs}, and construct the $\chi^2$ distribution:
\begin{eqnarray}
\chi^2\displaystyle &=&
\sum_{\rm bins} \quad \Bigl[\sum_{j,k \in ({B\to K^* \mu^+ \mu^- \,{\rm obs.}})} (O_j^{\rm exp} - O_j^{\rm th}) \, (\sigma^{({\rm bin})})^{-1}_{jk} \,  (O_k^{\rm exp} - O_k^{\rm th}) \Bigr] \nonumber\\
&& + \sum_{i\in ({\rm other \; obs.})} \frac{(O_i^{\rm exp} - O_i^{\rm th})^2}{(\sigma_i^{\rm exp})^2 + (\sigma_i^{\rm th})^2} 
\;,\label{eq:chi2abs}
\end{eqnarray}
where the central value of the experimental result and theoretical prediction of observable $i$ are given in $O_i^{\rm exp}$ and $O_i^{\rm th}$ respectively. The first term contains the contribution to the $\chi^2$ from
the $B\to K^* \mu^+ \mu^-$ observables. Here we include the experimental correlations among the $B\to K^* \mu^+ \mu^-$ observables, and the inverse of the correlation matrices $(\sigma^{({\rm bin})})^{-1}$ for each bin are taken from~\cite{1312.5267,Serra}. The second term is the standard $\chi^2$ without correlations for all the other observables. $\sigma_i^{\rm exp}$ and $\sigma_i^{\rm th}$ are the experimental and theoretical errors
respectively. 
Each MSSM point (defining a particular \emph{model}) is therefore assigned a $\chi^2$ value, which is directly related to the \emph{p-value} of the model: the probability that another measurement of these observables leads to worse agreement with the model predictions, assuming the model is true. All models with $p$-values greater than $\hat p$ lie inside a $(1-\hat p)\times 100$\% C.L. region. One-, two-, and three-sigma regions are defined as the 68.3\%, 95.4\% and 99.7\% C.L. regions respectively. 
Note that this approach is different from the one used in e.g. Refs.~\cite{1207.2753,1307.5683} as well as from the Bayesian procedure in Refs.~\cite{1111.1257,1205.1838,1310.2478} where the difference of the $\chi^2$ for each point with the $\chi^2$ of the best fit point is used to define the different allowed regions.

%%%%%%%%%%%%%%%%%%%%%%%%%%%%%%%%%%%%%%%%%%%%%%%%%%%%%%%%
\section{Results}
\label{sec:results}

\subsection{MSSM predictions for $B\to K^*\mu^+\mu^-$ observables}

\begin{figure}[!t]
\begin{center}
\includegraphics[width=5.5cm]{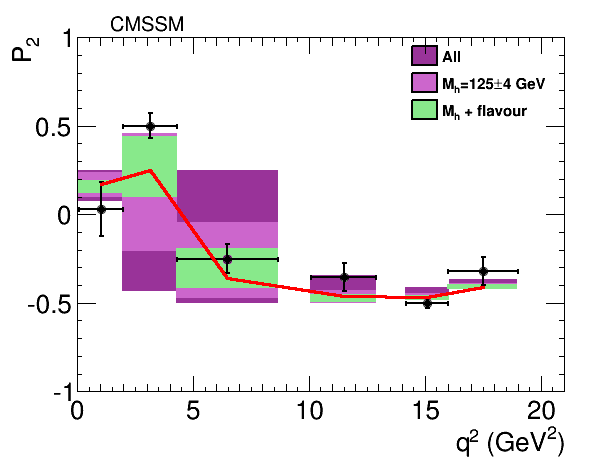}\includegraphics[width=5.5cm]{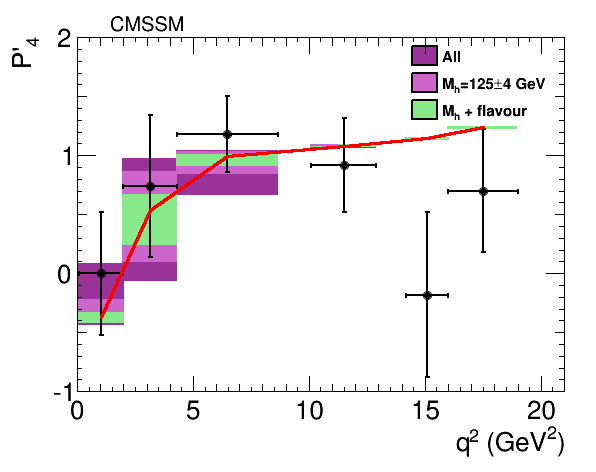}\includegraphics[width=5.5cm]{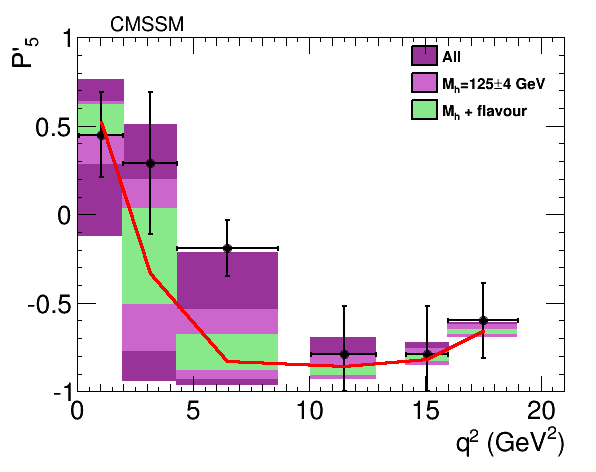}\\[5mm]
\includegraphics[width=5.5cm]{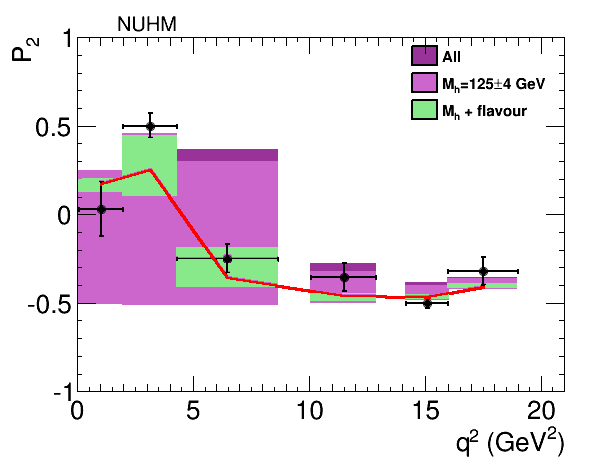}\includegraphics[width=5.5cm]{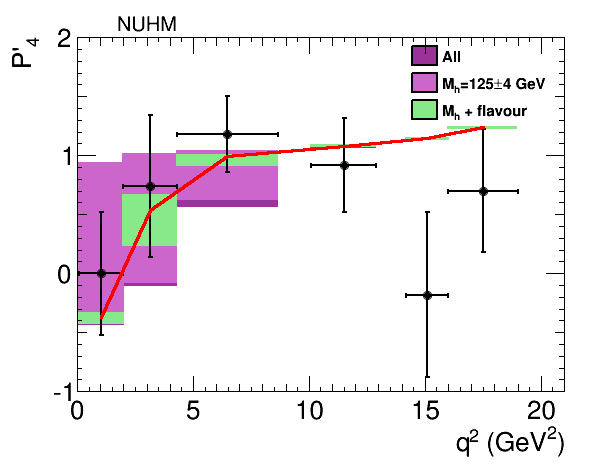}\includegraphics[width=5.5cm]{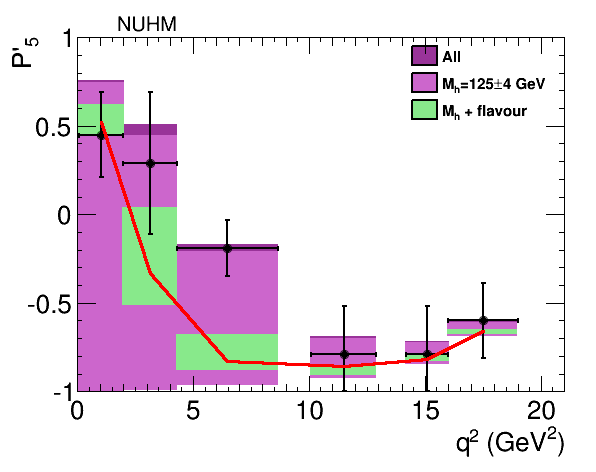}\\[5mm]
\includegraphics[width=5.5cm]{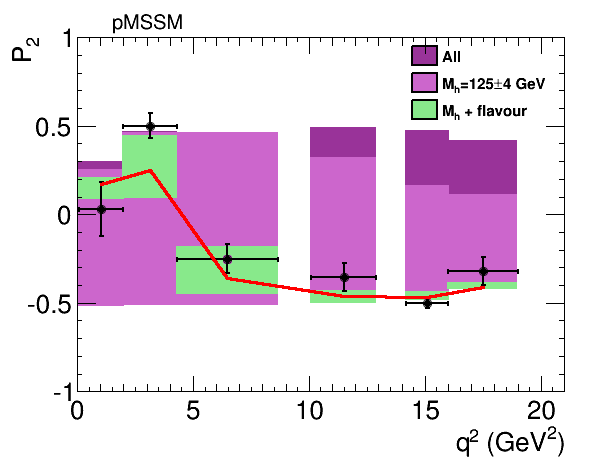}\includegraphics[width=5.5cm]{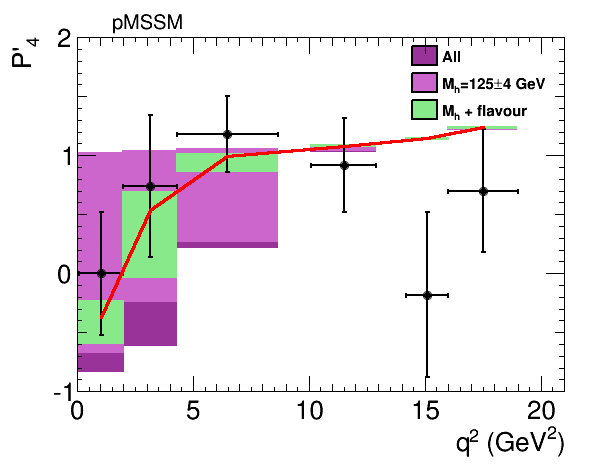}\includegraphics[width=5.5cm]{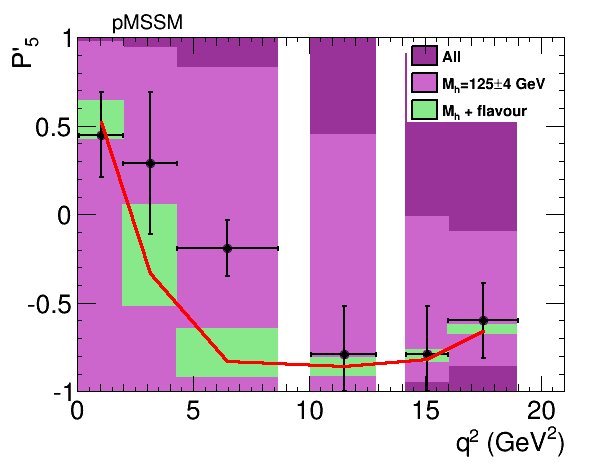}
\end{center}
\caption{MSSM predictions for $P_2$ (left), $P'_4$ (center) and $P'_5$ (right) within the CMSSM (upper row), NUHM (central row) and pMSSM (lower row). The dark purple regions correspond to the full predictions while the light purple zones are obtained after applying the Higgs mass constraints. The green bands include all flavour constraints (see Section \ref{sec:fit}).}
\label{fig:binned}
\end{figure}

We study first the reach of each MSSM scenario for all $B\to K^*\mu^+\mu^-$ observables, without applying any flavour or Higgs mass constraint. In this case the parameter space in each model is large enough to contribute significantly to each observable. The most interesting observables, $P_2$, $P'_4$ and $P'_5$, are shown in Fig.~\ref{fig:binned}, where the bands show the span of values allowed in this set-up. For comparison also the experimental results are shown (black data points), together with the SM central values (red line). As can be seen from the figure all MSSM scenarios can explain, within 1$\sigma$, each \emph{individual} tension for all observables\footnote{Note that $\av{P_4'}$ at low recoil cannot be changed by any significant amount since in this kinematic region this observable is independent of short distance physics, so that in particular the tension in the bin $[14.18,16]$ cannot be explained by New Physics.}. Any tension between the MSSM and $B\to K^*\mu^+\mu^-$ data can only come from 
correlations among this data, or from the combination with other constraints.

One such independent constraint is the Higgs mass. In order to see the effect of this constraint, we impose the condition $121 < M_h < 129$ GeV \cite{ATLAS:2013mma,CMS:xwa} on the scan points in each MSSM scenario. The MSSM contributions to $B\to K^*\mu^+\mu^-$ observables are slightly reduced, as can be seen in Fig.~\ref{fig:binned}, where the light purple bands show the span of values allowed from MSSM points satisfying the Higgs mass constraint. This reduction is inconsequential for all observables (again, barring possible correlations), except for $\av{P_5'}_{[4.3,8.68]}$, where one-sigma agreement is impossible in the constrained scenarios. The pMSSM on the other hand is general enough to provide solutions at the 1$\sigma$ level.

On the basis of these results, the relevant issue is how these observables are correlated in the different MSSM scenarios, among themselves and with other flavour observables. Even though Fig.~\ref{fig:binned} looks promising, the region in MSSM parameter space where a good agreement is found for e.g. $\av{P_2}_{[2,4.3]}$, could be predicting unacceptable values for $\av{P_5'}_{[2,4.3]}$, or for other observables such as BR$(B\to X_s\gamma)$. In this respect this is the main characteristic feature of a model-dependent analysis in contrast to a model-independent one where the various Wilson coefficients are independent parameters: correlations can lead to completely different global patterns even when the range of values for the Wilson coefficients are fixed.
\begin{figure}[!t]
\begin{center}
\includegraphics[width=5.5cm]{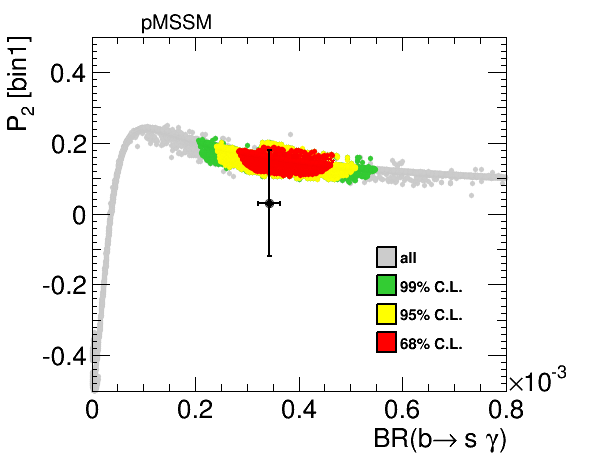}
\includegraphics[width=5.5cm]{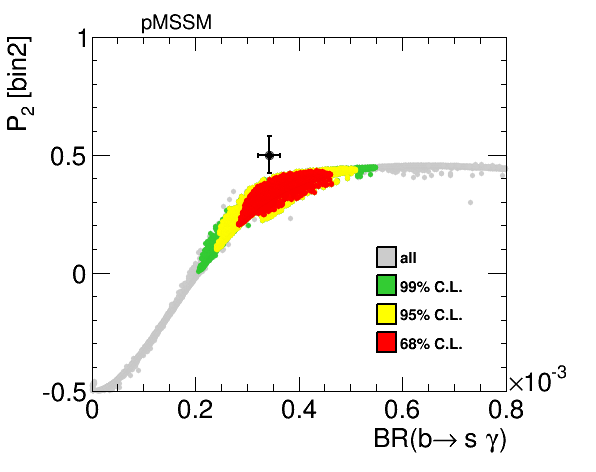}
\includegraphics[width=5.5cm]{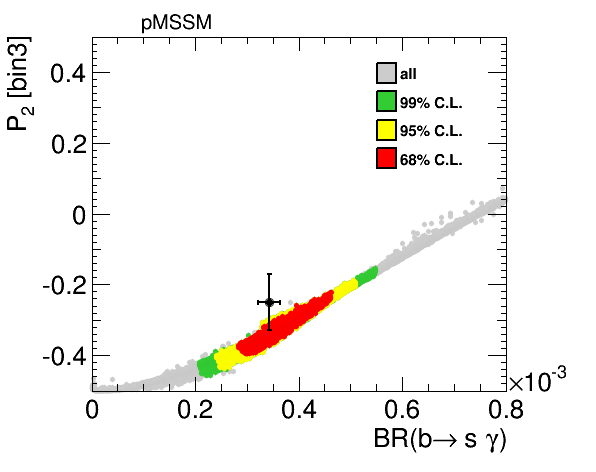}\\[2mm]
\includegraphics[width=5.5cm]{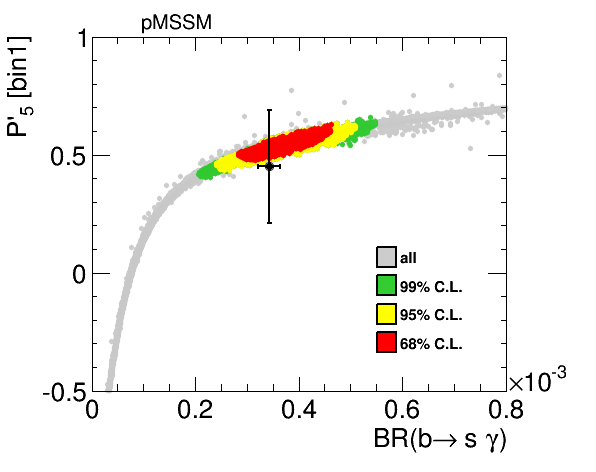}
\includegraphics[width=5.5cm]{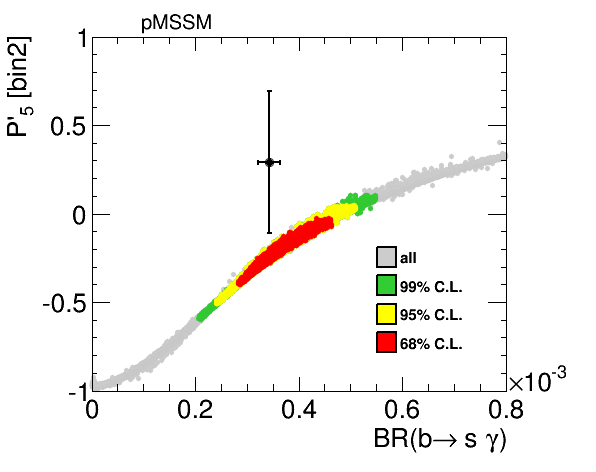}
\includegraphics[width=5.5cm]{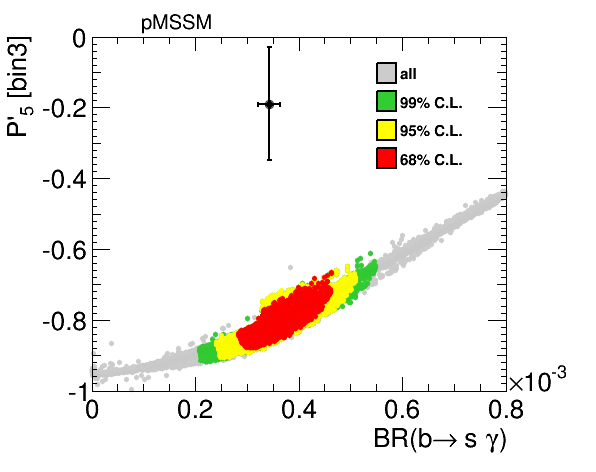}
\end{center}
\caption{Correlations among $P_2$ and $P'_5$ in the first three low $q^2$ bins and BR$(B\to X_s\gamma)$ in pMSSM. The black data points and error bars represent the experimental measurements.}
\label{fig:corrs_MSSM}
\end{figure}

We consider first several interesting examples of correlations among different observables within the MSSM. Even in a model-independent set-up, where all Wilson coefficients are treated as independent free parameters, strong correlations among different observables are present. In the MSSM, and in the case of small contributions to $\delta \C9$, sufficiently large values of $\delta \C7$ can enhance the relevant $B\to K^*\mu^+\mu^-$ observables enough to match the data. Irrespectively of whether these values are correlated to other Wilson coefficients, these large values of $\delta \C7$ are in tension with the measurement of BR$(B\to X_s\gamma)$. This is a (known) model-independent statement \cite{1207.2753,1307.5683}. 
With values for the Wilson coefficients reachable in the MSSM, the observables $\av{P_2}_{[2,4.3]}$, $\av{P_5'}_{[2,4.3]}$ and $\av{P_5'}_{[4.3,8.68]}$ can acquire large values in agreement with data; however this is mainly due to contributions to $\delta \C7$ that enhance BR$(B\to X_s\gamma)$ -- up to unacceptable levels in the case of $\av{P_5'}_{[4.3,8.68]}$ (see Fig.~\ref{fig:corrs_MSSM}). One has then to reach a statistical compromise between these two competing tensions, in order to decide whether $\delta \C7$ is more likely enhanced or not. Nevertheless, this tension will lead to a global enhancement of the $\chi^2$, and again the statistical framework will tell if this enhancement is admissible or not, in the latter case ruling out the corresponding MSSM scenario according to some statistical criteria. We should note however that both $B\to X_s\gamma$ and $B\to K^* \mu^-\mu^+$ observables show a preference for negative values of $\delta \C7$ (albeit to different degrees) which is by itself an 
intriguing fact.

In the model-independent framework, this tension can be cured in part with sizeable contributions to $\C9$, although it can never provide a good description of $\av{P_5'}_{[4.3,8.68]}$, at least without contributions to $\Cp{9}$ \cite{1307.5683}. In the MSSM, not only we have no means of generating any sizeable contribution to the coefficient $\Cp9$, but also any significant contribution to $\C9$ is correlated to contributions to other Wilson coefficients affecting the other observables. Some of these interesting patterns are shown in Fig.~\ref{fig:corrs_MSSM}.

\subsection{Constraints on the Wilson coefficients}
\label{sec:fit}

\begin{figure}[!t]
\begin{center}
\includegraphics[width=5.5cm]{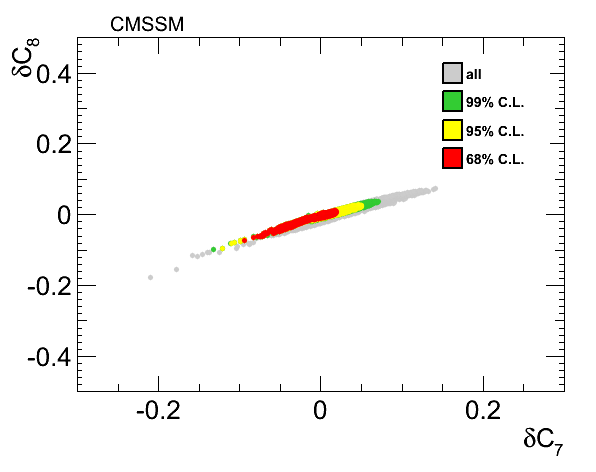}
\includegraphics[width=5.5cm]{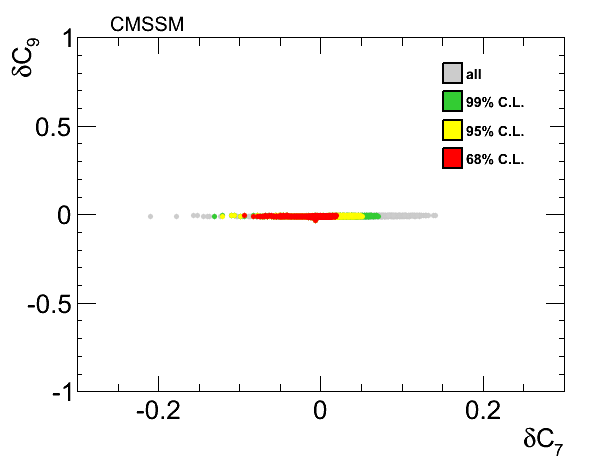}
\includegraphics[width=5.5cm]{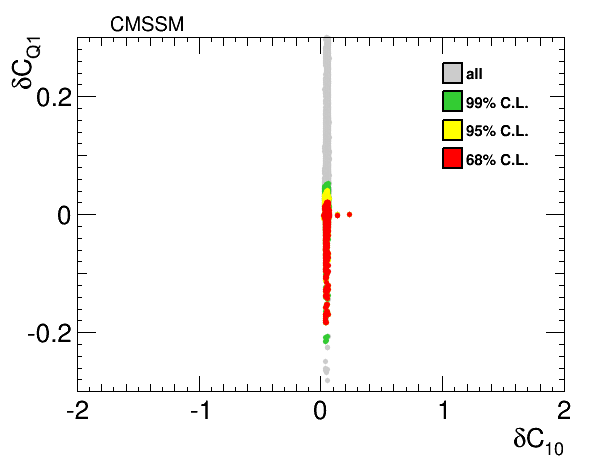}\\[5mm]
\includegraphics[width=5.5cm]{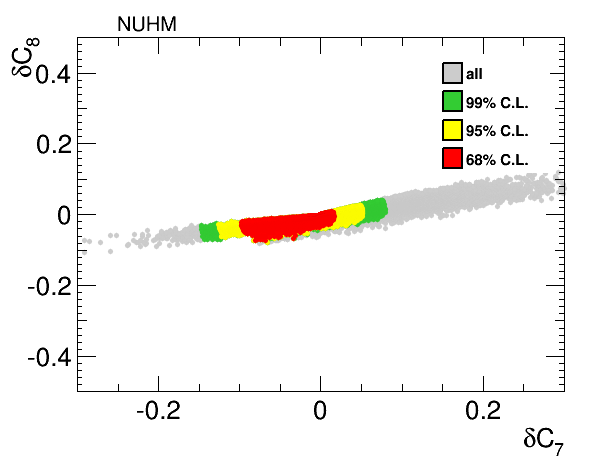}
\includegraphics[width=5.5cm]{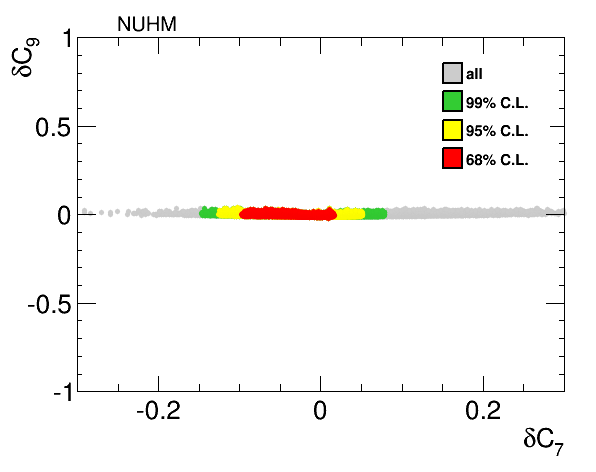}
\includegraphics[width=5.5cm]{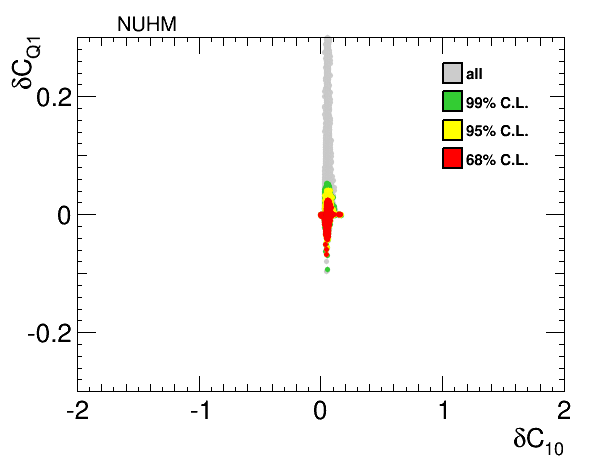}\\[5mm]
\includegraphics[width=5.5cm]{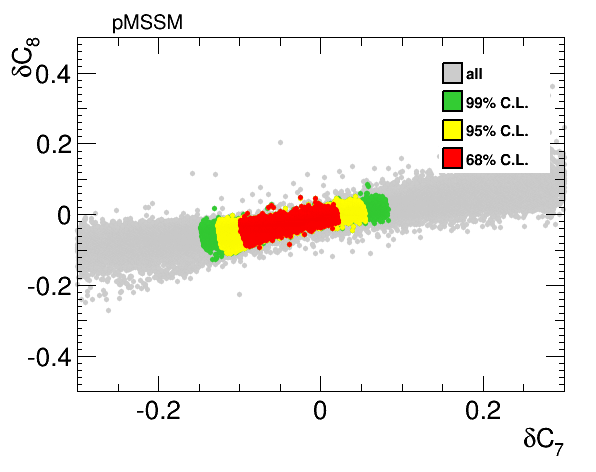}
\includegraphics[width=5.5cm]{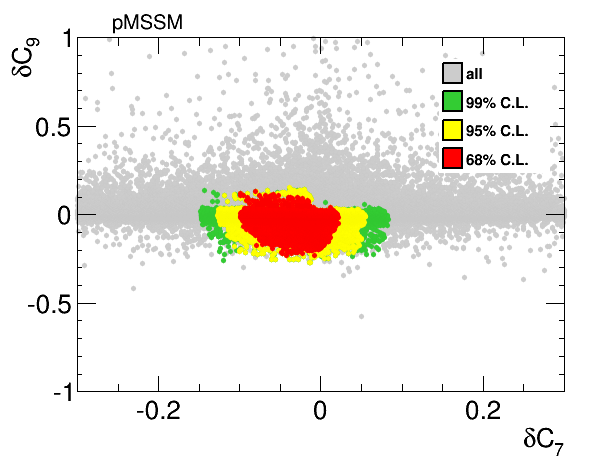}
\includegraphics[width=5.5cm]{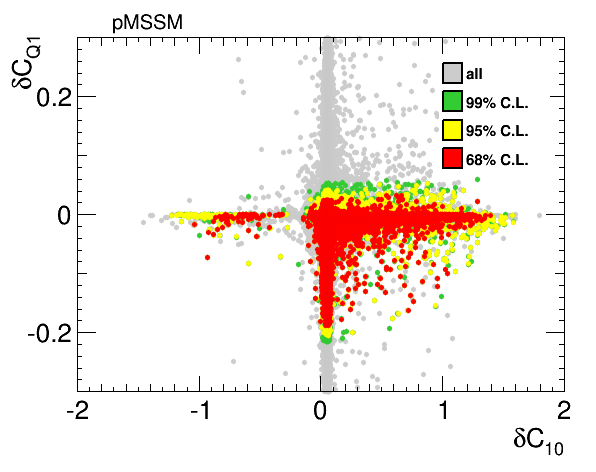}
\end{center}
\caption{Global fit to the NP coefficients $\delta \C{i}$ at the $\mu_b$ scale in CMSSM (upper row), in NUHM (central row) and in pMSSM (lower row), at 68\% C.L. (red) and 95\% C.L. (yellow) and 99\% C.L. (green) using the observables given in Table~\ref{tab:obs}, including the Higgs mass constraint.}
\label{fig:fit-mH}
\end{figure}

The flavour observables depend on the MSSM parameters only through the Wilson coefficients, and so is the case for the $\chi^2$ distribution. It is therefore quite convenient to look at the models and their C.L. regions in the space of Wilson coefficients. This is shown in Fig.~\ref{fig:fit-mH}, where all the points in the scans are shown (grey), indicating the values for the Wilson coefficients allowed within the MSSM, as well as the points satisfying the global 1,2,3$\,\sigma$ constraints (red, yellow, green). The Higgs mass constraint has also been imposed, all the points being consistent with a Higgs mass $121 < M_h < 129$ GeV.

The comparison with the model-independent fit in Ref.~\cite{1307.5683} as well as with Refs.~\cite{1308.1501,1310.2478} is not straightforward. In particular, the results in Fig.~\ref{fig:fit-mH} do not contradict the result of Ref.~\cite{1307.5683} that a scenario with $\delta \C9\sim -1.2$ and $\delta \C7\sim -0.1$ is the most favourable one and with respect to which the SM hypothesis has a pull of $\sim 3$-$4\,\sigma$. Notice that: (a) No such large values for $\delta \C9$ are allowed within these MSSM scenarios, and (b) even though the $\chi^2$ grows for $\delta \C9$ decreasing from $\sim -0.2$ (at least within the limited number of scan points, see the center plot in the last row of Fig.~\ref{fig:fit-mH}), these points lead simultaneously to values for other Wilson coefficients away from zero, and therefore cannot be compared with a model-independent analysis where $\delta \C9$ can be decreased independently of the values of the other coefficients. As mentioned earlier, this is the result of strong 
model-dependent correlations 
absent within the model-independent approach.

As a conclusion, the most favourable MSSM models given the $B\to K^*\mu^+\mu^-$ data lead only to small deviations for the Wilson coefficients with respect to their SM values. The best scenarios in terms of Wilson coefficients, known from model-independent analyses, are not reachable within the constrained MSSM models, or even within the more generous pMSSM. However, the overall agreement is fairly good, with regions in SUSY parameter space where the absolute $\chi^2$ is sufficiently small. Within this statistical criteria the MSSM is still well compatible with the data. 

It is instructive to go back to Fig.~\ref{fig:binned} and impose the constraints derived from the global fit. This is shown as the over-imposed green bands, showing the span of models that survive the 1$\,\sigma$ constraints from the global fit. We see that good agreement with $\av{P_2}_{[2,4.3]}$ and $\av{P'_5}_{[2,4.3]}$ can be obtained, improving the situation with respect to the SM. On the other hand, the tension in $\av{P'_5}_{[4.3,8.68]}$ cannot be explained within the pMSSM. This is compatible with the previous observation that, without significant contributions to $\C9$, a large value for $\av{P'_5}_{[4.3,8.68]}$ requires extreme values for other Wilson coefficients that, while reachable within the MSSM, are ruled out by other flavour bounds.

\subsection{Implications for MSSM parameters}

The results of the global fit can be used to put constraints on the supersymmetric parameters. This is done by computing such quantities from the input parameters within each MSSM scenario which satisfies the global constraints in the fit. 

\begin{figure}[!t]
\begin{center}
\includegraphics[width=7.5cm]{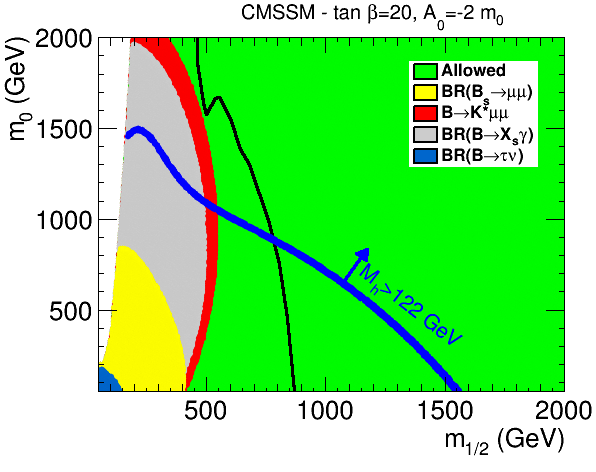}\hspace{1cm}
\includegraphics[width=7.5cm]{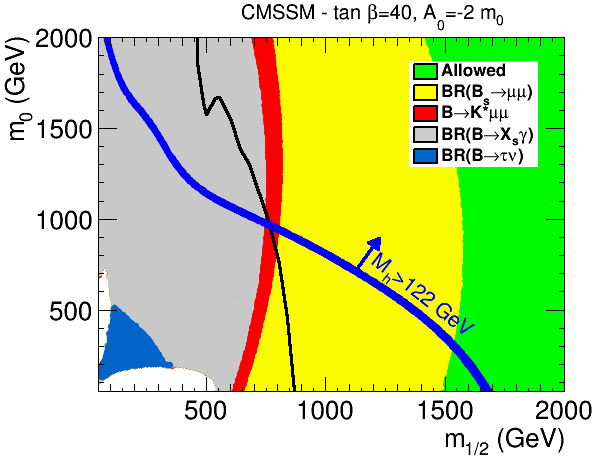}
\end{center}
\caption{Flavour constraints in CMSSM in the plane ($m_{1/2},m_0$) for $A_0=-2 m_0$ and $\tan\beta =20$ (left) and 40 (right). The black line represents the experimental direct SUSY search limit with 20.3~fb$^{-1}$ of data at 8 TeV \cite{ATLAS:2013fha}.}
\label{fig:cmssm}
\end{figure}
\begin{figure}[!t]
\begin{center}
\includegraphics[width=7.5cm]{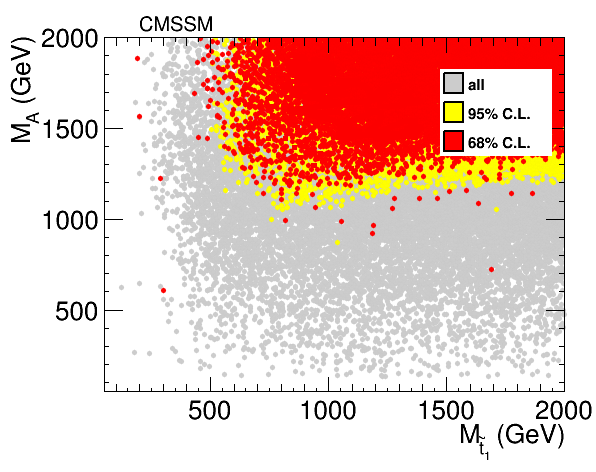}\hspace{1cm}
\includegraphics[width=7.5cm]{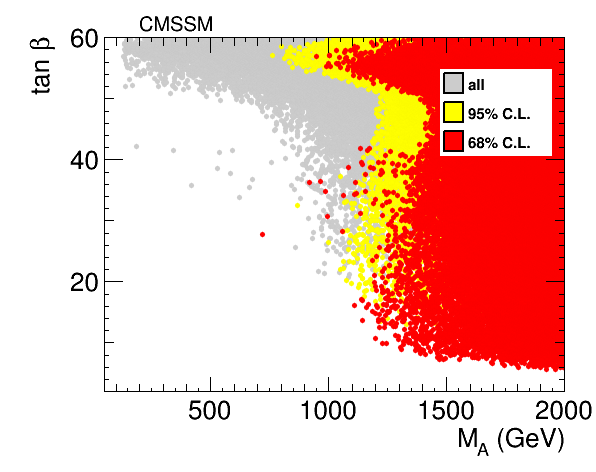}\\
\includegraphics[width=7.5cm]{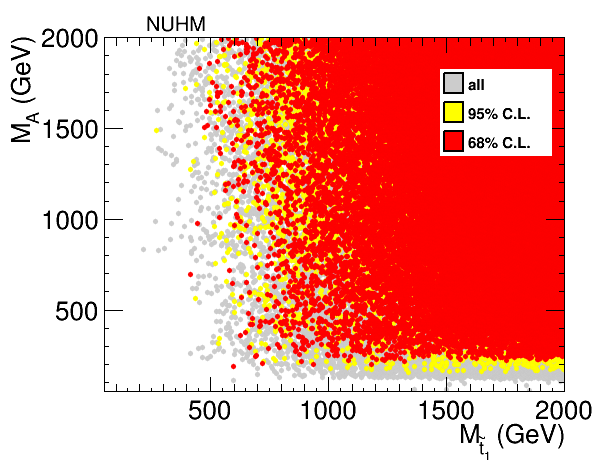}\hspace{1cm}
\includegraphics[width=7.5cm]{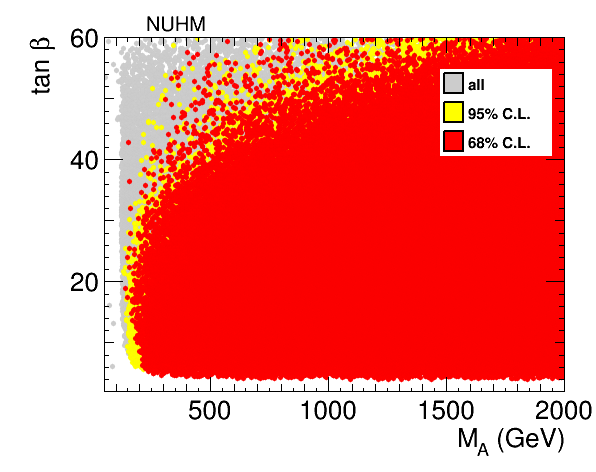}\\
\includegraphics[width=7.5cm]{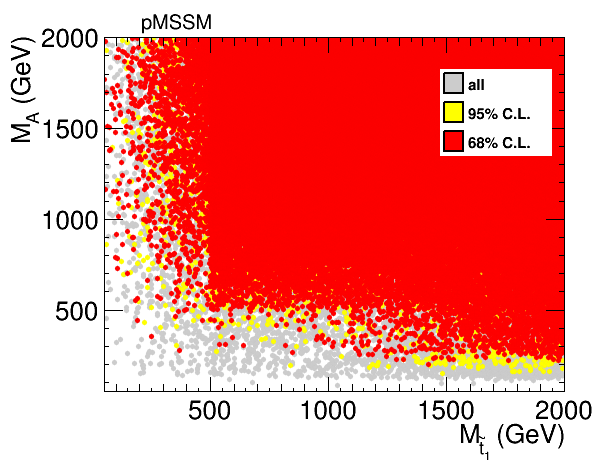}\hspace{1cm}
\includegraphics[width=7.5cm]{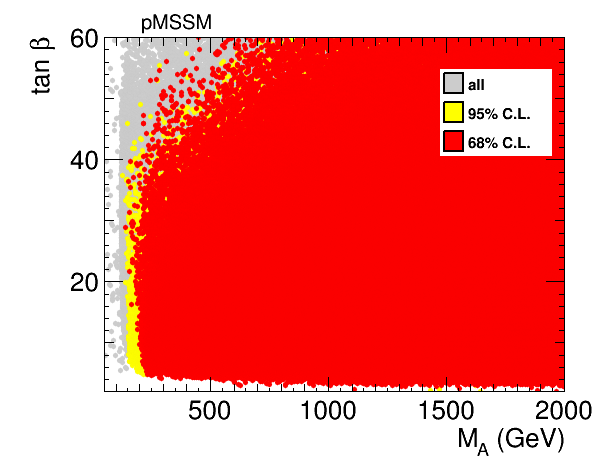}
\end{center}
\caption{Allowed regions at 68\% C.L. (red) and 95\% C.L. (yellow) in the planes $(M_{\tilde{t}_1},M_A)$  (left) and $(M_A,\tan\beta)$ (right) in CMSSM (upper row), in NUHM (central row) and in pMSSM (lower row).}
\label{fig:excl2d}
\end{figure}

First, in order to see the impact of the newly measured $B\to K^*\mu^+\mu^-$ observables in a simple framework, in Fig.~\ref{fig:cmssm}
we compare the constraining power 
of the fit to $B\to K^*\mu^+\mu^-$ observables with some of the other flavour observables ({\it i.e.} the Branching Ratios of $B_s \to \mu^+ \mu^-$, $B\to X_s\gamma$ 
and $B\to \tau \nu$) in the CMSSM. 
We have kept $\tan \beta$ to the fixed values of 20 and 40 and the trilinear soft breaking 
parameter $A_0=-2 m_0$ (this choice allows us to have a better agreement with the Higgs mass). In this way, the results in the two dimensional ($m_{1/2},m_0$) plane can be easily understood.
The constraints from the combination of $B\to K^*\mu^+\mu^-$ observables are shown in red and can be compared to other flavour constraints shown in different colours according to the legends in the plots. It is remarkable that the constraints from $B\to K^*\mu^+\mu^-$ using the recent measurements are at the same level as the ones from the well-known BR($B\to X_s \gamma$) constraints. Also, the constraints from $B_s\to \mu^+\mu^-$ while being very strong at large $\tan\beta$, are less restrictive 
for lower values of $\tan\beta$ where the $B\to K^*\mu^+\mu^-$ constraints dominate. For comparison we show also in the figure the direct SUSY search limits in the same plane from ATLAS with 20.3 fb$^{-1}$ of data at $\sqrt{s}=8$ TeV \cite{ATLAS:2013fha}, as well as the region where the Higgs mass is above 122 GeV. It is also very interesting to notice that the flavour constraints start being stronger than the direct search limits for large values of $\tan\beta \gtrsim 20$.

Next we consider the MSSM scenarios in full generality without fixing any parameters. The constraints from the global fit on the physical parameters $M_{\tilde{t}_1}$, $M_A$ and $\tan\beta$ are shown in  Fig.~\ref{fig:excl2d}. Several interesting features can be observed. First, while the constraints are very strong in the CMSSM, they can be easily relaxed in the more general scenarios NUHM or pMSSM. In the case of the CMSSM, the pseudo-scalar Higgs mass is constrained quite generally to be above $\sim 1$ TeV.
This strong limit from flavour physics is independent of the value of $\tan\beta$. Another interesting feature is the impact on the lightest stop mass. In the constrained scenarios, flavour constraints require the lightest stop to be heavier than $\sim 500$ GeV. On the other hand, in the pMSSM the flavour constraints are much weaker, excluding only scenarios where  $M_A$ and $M_{\tilde{t}_1}$ are both small, or with light $M_A$ and large $\tan\beta$.

\begin{figure}[!t]
\begin{center}
\includegraphics[width=5.5cm]{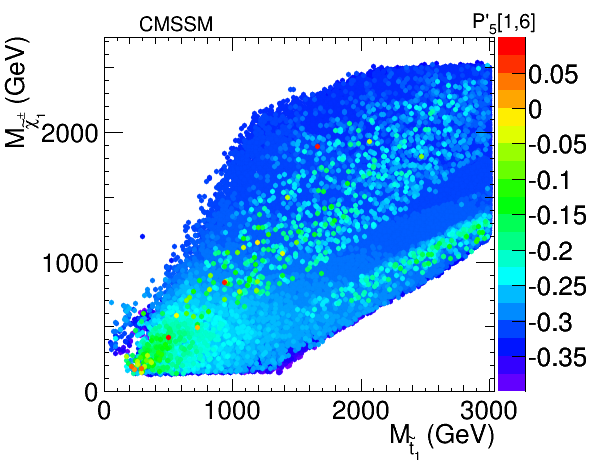}
\includegraphics[width=5.5cm]{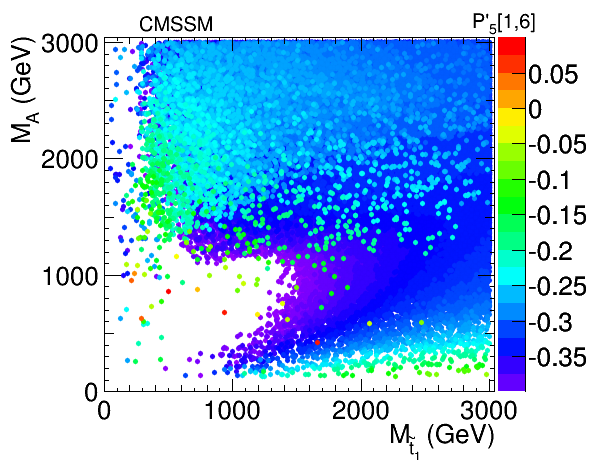}
\includegraphics[width=5.5cm]{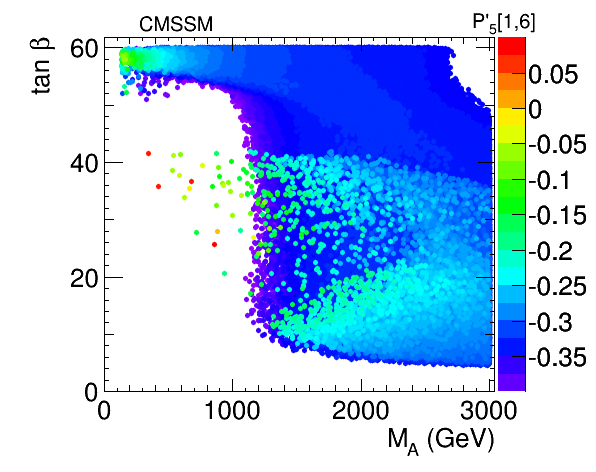}\\[5mm]
\includegraphics[width=5.5cm]{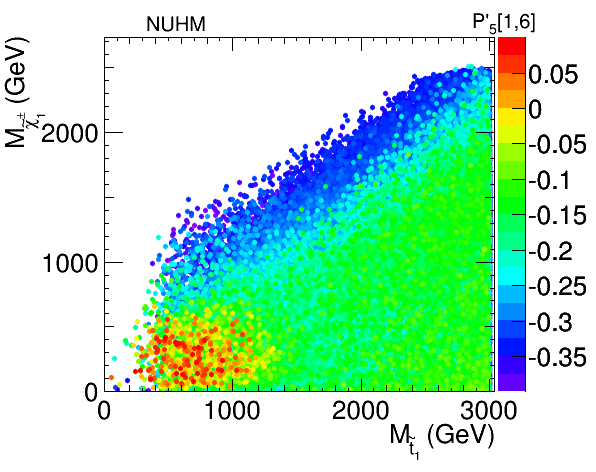}
\includegraphics[width=5.5cm]{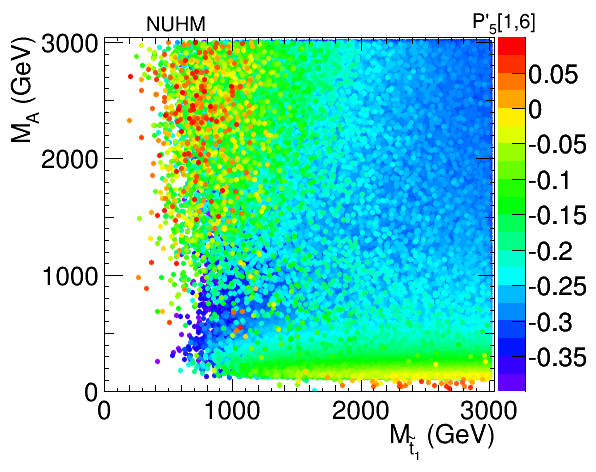}
\includegraphics[width=5.5cm]{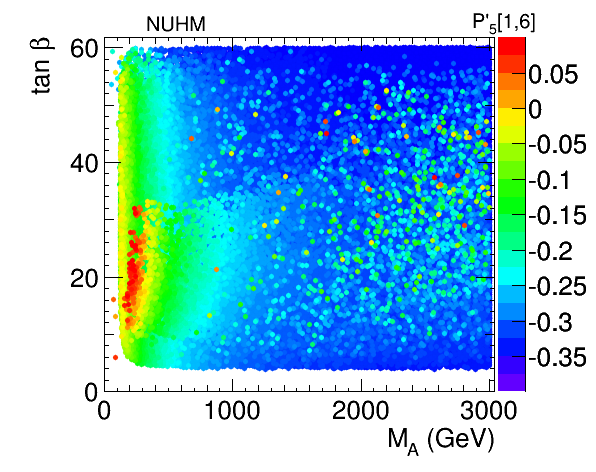}\\[5mm]
\includegraphics[width=5.5cm]{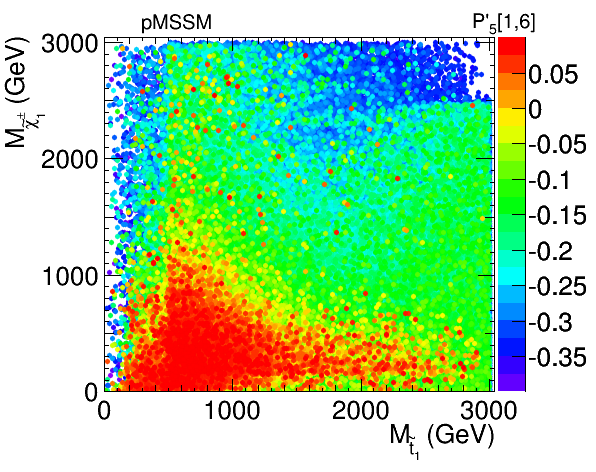}
\includegraphics[width=5.5cm]{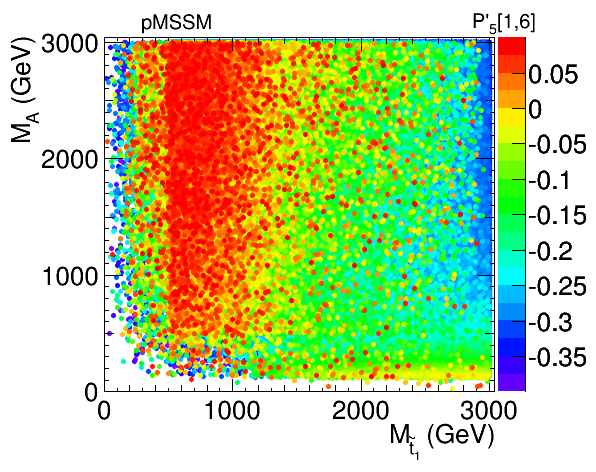}
\includegraphics[width=5.5cm]{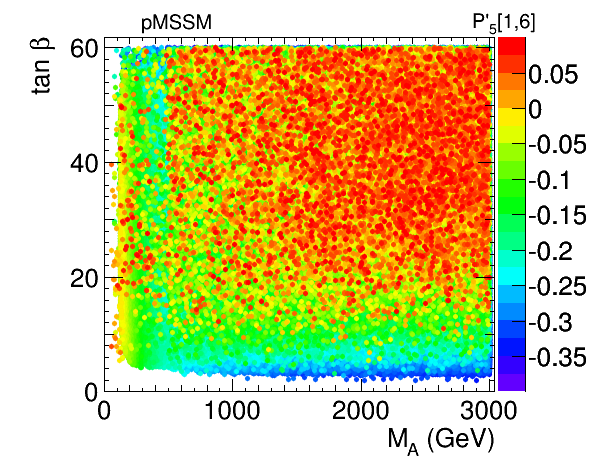}
\end{center}
\caption{Intensity of $P'_5$ in the $[1, 6]$ GeV bin in colour scale, in the planes $(M_{\tilde{t}_1},M_{\tilde{\chi}^\pm_1})$ (left), $(M_{\tilde{t}_1},M_A)$ (centre) and $(M_A,\tan\beta)$ (right) in CMSSM (upper row), in NUHM (central row) and in pMSSM (lower row). The SM prediction for $P'_5$ in the $[1, 6]$ GeV bin is $-0.34 \pm 0.10$, and the experimental result is $0.21 \pm 0.21 \pm 0.03$.}
\label{fig:P5-3d}
\end{figure}

Finally, we devote a special attention to $P'_5$, as this is the only observable presenting more than 2$\sigma$ tension in the MSSM, and study how the value of the observable $\av{P_5'}_{[1,6]}$ changes as a function of the MSSM parameters. The results are displayed in Fig.~\ref{fig:P5-3d}.
As can be seen, the tension with respect to the experimental result is reduced for small chargino or stop masses, while the other SUSY parameters are less relevant.

%%%%%%%%%%%%%%%%%%%%%%%%%%%%%%%%%%%%%%%%%%%%%%%%%%%%%%%%
\section{Conclusions}
\label{sec:conclusions}

The angular distribution of the decay $B\to K^*\mu^+\mu^-$ provides a large number of observables, probing NP in a complementary way to other related flavour processes.
This exclusive mode is under good theoretical control, with the main hadronic uncertainties stemming from the form factors and (at large recoil) unknown power corrections. In this context the use of \emph{optimised} observables such as the set \{$P_i^{(\prime)}$\} is very convenient in phenomenological analyses.

Recent LHCb measurements of these  optimised observables have shown certain discrepancies with SM predictions, specifically in some of the $P_2$ and $P_5^\prime$ bins. Model-independent global fits for the Wilson coefficients have shown that the 
tensions can be relaxed if $\delta {\cal C}_9 \sim -1$.
In this paper we have studied the implications of these new results along with other relevant $b\to s$ transitions on several MSSM scenarios: the constrained CMSSM and NUHM scenarios and also the more general pMSSM framework.
While large negative contributions to ${\cal C}_9$ are not accessible within the MSSM, 
it is interesting to note that the MSSM has the potential to reproduce each of the measurements individually at the 1$\,\sigma$ level. Correlations among the different observables are then the central issue.

In order to study the impact of the recent LHCb results, we have first performed a global fit to
the Wilson coefficients, including most of the available $b\to s$ modes.
The comparison of this analysis with previous model-independent analyses is not straightforward, but can be understood in terms of model-dependent correlations and a different statistical set-up.
In our analysis, the best fit values for the Wilson coefficients not only depend on the correlations 
among the different flavour observables but also on the correlations between the various Wilson coefficients in each specific MSSM scenario.
Even though the small value of ${\cal C}_9\sim 3$ favourable from model-independent analyses is not accessible within MSSM,
there are regions of the MSSM parameter space that (within some statistical criteria) are still well compatible with the experimental data. 

We have translated the constraints on different SUSY parameters in the three aforementioned scenarios.
These constraints provide complementary information to direct searches, and show that the data on $B\to K^*\mu^+\mu^-$ has become competitive with the traditional modes such as $B\to X_s\gamma$ and $B_s\to \mu^+\mu^-$.

The observable $P_5^\prime$ remains nevertheless a challenge for SUSY models. We have shown its dependence
on some of the more relevant SUSY parameters, making it possible to easily look for the 
preferred region of parameter space within the mentioned MSSM scenarios when experimental updates on $P_5^\prime$ are available. We are looking forward to the experimental update including the full 3~fb$^{-1}$ data-set collected at LHCb.

\section*{Acknowledgements}

We would like to thank J.~Matias for valuable contributions concerning the SM predictions of $B\to K^*\mu^+\mu^-$ observables, and T. Hurth for innumerable useful discussions. J.V. is funded by the Deutsche Forschungsgemeinschaft (DFG) within research 
unit FOR 1873 (QFET).

%%%%%%%%%%%%%%%%%%%%%%%%%%%%%%%%%%%%%%%%%%%%%%%%%%%%%%%%

\end{document}